\documentclass{ws-ijmpd}
\usepackage[super,compress]{cite}
\usepackage[hidelinks]{hyperref}
\usepackage[caption = false]{subfig}
\usepackage{tikz}
\usetikzlibrary{patterns}

\graphicspath{{.}{../Plots/}}
\begin{document}

\markboth{Justin L. Ripley}
{Numerical methods for Horndeski gravity}

%
\catchline{}{}{}{}{}
%

\title{NUMERICAL RELATIVITY FOR HORNDESKI GRAVITY}

\author{JUSTIN L. RIPLEY}

\address{Department of Applied Mathematics and Theoretical Physics, \\
University of Cambridge, Cambridge, CB3 0WA,
United Kingdom
\\
\;
\\
Illinois Center for Advanced Studies of the Universe \& Department of Physics,
University of Illinois at Urbana-Champaign, Urbana, Illinois 61801, USA
\\
\;
\\
ripley@illinois.edu}

\maketitle

\begin{history}
\received{Day Month Year}
\revised{Day Month Year}
\end{history}

\begin{abstract}
   We present an overview of recent developments in the numerical solution
   of Horndeski gravity theories, which are the class of all scalar-tensor
   theories of gravity that have second order equations of motion.
   We review several methods that have been used to establish
   well-posed initial value problems for these theories,
   and discuss well-posed formulations of the constraint equations.
   We also discuss global aspects of exact, strongly coupled solutions to 
   some of Horndeski gravity theories: 
   the formation of shocks, the loss of hyperbolicity, 
   and the formation of naked curvature singularities.
   Finally we discuss numerical solutions to binary black hole and
   neutron star systems for several Horndeski theories.
\end{abstract}


\ccode{PACS numbers:}

\tableofcontents

\section{Introduction	
   \label{sec:introduction}
}
Due to the complicated structure of the Einstein equations,
understanding the merger of black holes and neutron stars--where 
gravity is in the
\emph{strong field, dynamical regime}--requires the use of
computers to solve the equations of motion\cite{alcubierre2008introduction,
baumgarte2010numerical}. 
A key technical step that led to the first stable numerical
simulations of binary black hole systems in General Relativity (GR)
was the construction of well-posed formulations of the  Einstein 
equations\cite{Pretorius:2005gq,
Campanelli:2005dd,Baker:2005vv,Pretorius:2007nq}.
Since then, the observation of binary black holes (and neutron stars)
by the LIGO/Virgo/KAGRA
collaboration\cite{LIGOScientific:2016aoc,
LIGOScientific:2016sjg,LIGOScientific:2017vwq} 
has spurred interest in understanding how modifications to GR
could affect the classical dynamics of those objects\cite{Yunes:2013dva,
Berti:2015itd,Foucart:2022iwu}. 
As of the writing of this review,
the gravitational wave observations of binary black holes and neutron
stars by the LIGO/Virgo/KAGRA collaboration have so far
been found to be consistent with the predictions of 
GR\cite{Yunes:2013dva,LIGOScientific:2016lio,LIGOScientific:2019fpa,
LIGOScientific:2020tif,Krishnendu:2021fga}.
Despite this,  
as model-dependent tests of GR can allow for more precise tests of GR,
there has been an extensive effort by many researchers
to model the dynamics of binary black holes
and neutron stars in \emph{modified} (non-GR) 
theories of gravity\cite{Berti:2015itd,Foucart:2022iwu}.

The Horndeski theories\cite{horndeski_paper,Deffayet:2009wt,Deffayet:2011gz},
which encompass all theories of gravity
that have a scalar and tensor field and that have second 
order equations of motion, have attracted particular attention
as they allow for scalar ``hairy'' black hole and neutron
star solutions\cite{Damour:1993hw,Kanti:1995vq,Sotiriou:2013qea},
and have been used to model dark energy\cite{Copeland:2006wr}\footnote{We note
though that many Horndeski models that have been used to model energy
are have been highly constrained by recent gravitational 
wave measurements; for one example constraint see
\refcite{Baker:2017hug}}.
As was the case for GR around 25 years ago, 
a major challenge to numerically solving the equations
of motion for Horndeski (and other modified) theories 
of gravity has been in constructing 
well-posed formulations of their equations of motion, that also
allow for numerically stable evolution of black holes 
and neutron stars\cite{Cayuso:2017iqc,
Allwright:2018rut,Bernard:2019fjb,Kovacs:2019jqj,
Witek:2020uzz,Kovacs:2020pns,Kovacs:2020ywu}.
An additional challenge to constructing numerical solutions
in these theories has been in determining the properties of 
\emph{global} solutions to Horndeski equations.
Despite these challenges, recent mathematical and numerical
relativity work indicates that it should be 
possible to simulate binary black hole/neutron star 
mergers for most, if not all, of the Horndeski gravity theories of
astrophysical and cosmological interest.

In this review, we summarize recent work on the local and global
properties of solutions to
the Horndeski gravity theories, along with numerical
relativity studies of the Horndeski gravity theories in the
context of binary black hole/neutron star systems.
Here we are purely concerned with the classical evolution
of Horndeski gravity theories.
We do not discuss
the quantum mechanical viability of any particular Horndeski theory,
although we note that as these theories have second order equations
of motion, naive quantization of the theories
do not suffer from the 
\emph{Ostrogradski instability}\cite{Woodard:2015zca,Kobayashi:2019hrl}.
We do not provide a detailed review of
the current observational and
theoretical constraints on the Horndeski gravity theories, nor
do we discuss other methods to construct solutions to the
Horndeski gravity theories, such as Post-Newtonian methods.
Finally, we do not discuss recent generalizations of the Horndeski
gravity theories\cite{Kobayashi:2019hrl}.

A brief summary of what we cover:

We first discuss a generic perturbative
\emph{order-reduction scheme}, which when applied to the Horndeski
equations of motion, always leads to a strongly hyperbolic system of
evolution equations and to elliptic constraint equations, with
no restriction of the nature of the solutions 
(Sec.~\ref{sec:perturbative_well_posedness}).
The disadvantage of this approach is the \emph{secular growth}
of errors which generically occur in higher order perturbative solutions.
These may be addressed via the 
\emph{numerical dynamical renormalization group}
(Sec.~\ref{sec:perturbative_secular_instabilities}), although
that technique has yet to be applied to the Horndeski gravity theories.

We next turn to \emph{direct} approaches to solving the Horndeski
equations of motion, that is approaches that solve the complete
Horndeski equations of motion without approximation.
Despite the considerable complexity of the Horndeski equations
of motion (see \ref{sec:eom_horndeski_gravity}), 
their general structure as partial differential equations (PDE) 
are remarkably similar to the Einstein 
equations\cite{Kovacs:2020pns,Kovacs:2020ywu,
Kovacs:2021lgk,Kovacs:2021vdk,Reall:2021voz}. 
For \emph{weakly-coupled} solutions (Sec.~\ref{sec:weak_coupling}),
the equations of motion can be cast
into a strongly hyperbolic form (Sec.~\ref{sec:local_well_posedness_exact_eom}),
and the constraint equations can be recast as a set of elliptic
partial differential equations (Sec.~\ref{sec:constructing_initial_data}).
For \emph{strongly-coupled} solutions, some of the Horndeski theories 
can break down in a way  
not seen in solutions to the Einstein equations:
shocks can form from smooth initial data\footnote{Albeit shocks can 
form from smooth initial data 
for more ``standard'' matter fields, such
as for perfect fluids\cite{rezzolla2013relativistic}.},
the equations of motion can lose their hyperbolic properties 
in \emph{elliptic regions}, and curvature singularities
may form (Sec.~\ref{sec:failure_global}).
Elliptic regions and curvature singularities can form \emph{outside}
of event horizons for some Horndeski theories.
As both indicate a loss of predictivity, those theories
have solutions that violate 
weak cosmic censorship\cite{Ripley:2019hxt,Ripley:2019irj,
Figueras:2020dzx,East:2021bqk} (the formation of elliptic
regions and/or curvature singularities 
could potentially be prevented by \emph{fixing} the equations
of motion, although more work remains to show this can always be done; 
see Sec.~\ref{sec:fixing_eqns}).
This being said, the predominant view of Horndeski theories is that they
should be viewed as parametrizing \emph{effective} deviations from
the Einstein equations. 
With this point of view, the breakdown of a Horndeski
theory in a given solution can be interpreted as that solution lying outside
the regime of applicability of the theory, 
and not that it is invalid to ever make use of that theory.

We present the full equations of motion for Horndeski gravity
in \ref{sec:eom_horndeski_gravity}. 
We provide a short review of the basic concepts of
well-posedness of evolution equations in
\ref{sec:review_hyperbolicity},
and of mixed-type partial differential equation in
\ref{sec:mixed_type}.
We discuss some more general properties
of the principal symbol of the Horndeski gravity theories
in \ref{sec:general_properties_principal_symbol}, and collect a few
identities that are useful for deriving the constraint equations of the
Horndeski gravity theories in \ref{sec:spatial_conformal_decomposition}.
We review an effective field
theory-styled ``derivation'' of Einstein scalar Gauss-Bonnet
gravity (a Horndeski gravity theories that has attracted particular
attention because it admits scalar hairy black hole solutions) in 
\ref{sec:4dst_gradient_expansion}.

Our notation is: the metric has $-+++$ signature, 
lower-case Greek letters index spacetime tensor components,
lower-case Latin letter index spatial tensor components,
the Riemann tensor is 
$R^{\alpha}{}_{\mu\beta\nu}
=
\partial_{\beta}\Gamma^{\alpha}_{\mu\nu}-\cdots$,
``$l.o.t.$'' means ``lower order terms'' in derivatives 
(for example $\partial_t^2\phi+\partial_t\phi = \partial_t^2\phi+l.o.t.$),
and $\mathtt{P}\left[\cdots\right]$ means take the principal part of 
a set of equations, 
(for example 
$\mathtt{P}\left[\partial_t^2\phi+\partial_t\phi\right] = \partial_t^2\phi$).
We set $8\pi G = c = 1$.

\section{Horndeski gravity
   \label{sec:horndeski_gravity}
}	

The Horndeski gravity theories consist of all classical field theories
that have a tensor field $g_{\mu\nu}$, a scalar field $\phi$, 
have only up to second order derivatives in the action,
and which have second-order equations of 
motion\cite{horndeski_paper,Deffayet:2009wt,Deffayet:2011gz}.
We write the Horndeski action as
\begin{align}
\label{eq:general_horndeski_action}
   S
   =
   \int d^4x\sqrt{-g}
   \left(
      \mathcal{L}_1
      +
      \mathcal{L}_2
      +
      \mathcal{L}_3
      +
      \mathcal{L}_4
      +
      \mathcal{L}_5
   \right)
   ,
\end{align}
where
\begin{align}
   \mathcal{L}_1
   \equiv&
   \frac{1}{2}R
   +
   X
   -
   V\left(\phi\right)
   ,\\
   \mathcal{L}_2
   \equiv&
   \mathcal{G}_2\left(\phi,X\right)
   ,\\
   \mathcal{L}_3
   \equiv&
   \mathcal{G}_3\left(\phi,X\right)\Box\phi
   ,\\
   \mathcal{L}_4
   \equiv&
   \mathcal{G}_4\left(\phi,X\right)R
   +
   \partial_X\mathcal{G}_4\left(\phi,X\right)
   \delta^{\gamma_1\gamma_2}_{\delta_1\delta_2}
   \nabla_{\gamma_1}\nabla^{\delta_1}\phi
   \nabla_{\gamma_2}\nabla^{\delta_2}\phi
   ,\\
   \mathcal{L}_5
   \equiv&
   \mathcal{G}_5\left(\phi,X\right)G_{\mu\nu}\nabla^{\mu}\nabla^{\nu}\phi
   -
   \frac{1}{6}\partial_X\mathcal{G}_5\left(\phi,X\right)
   \delta^{\gamma_1\gamma_2\gamma_3}_{\delta_1\delta_2\delta_3}
   \nabla_{\gamma_1}\nabla^{\delta_1}\phi
   \nabla_{\gamma_2}\nabla^{\delta_2}\phi
   \nabla_{\gamma_3}\nabla^{\delta_3}\phi
   ,
\end{align}
$\delta^{\cdots}_{\cdots}$ corresponds to the (generalized) Kronecker delta,
and
\begin{align}
   X
   \equiv
   -
   \frac{1}{2}g^{\mu\nu}\nabla_{\mu}\phi\nabla_{\nu}\phi
   .
\end{align}
Following Papallo and Reall\cite{Papallo:2017qvl}, we assume
\begin{align}
   \label{eq:coefficient_assumptions}
   \mathcal{G}_2\left(\phi,0\right)
   =
   \partial_X\mathcal{G}_2\left(\phi,0\right)
   =
   \mathcal{G}_3\left(\phi,0\right)
   =
   \mathcal{G}_4\left(0,0\right)
   =
   \mathcal{G}_5\left(0,0\right)
   =
   0
   .
\end{align}
to remove degeneracies between the various $\mathcal{L}$ terms
through field redefinitions $\phi\to f(\phi)$.
The equations of motion are given by
\begin{align}
   E^{(g)}_{\mu\nu}
   &\equiv
   -
   \frac{1}{\sqrt{-g}}\frac{\delta S}{\delta g^{\mu\nu}}
   =
   0
   ,\\
   E^{(\phi)}
   &\equiv
   -
   \frac{1}{\sqrt{-g}}\frac{\delta S}{\delta\phi} 
   =
   0
   .
\end{align}
We state the full equations of motion in \ref{sec:eom_horndeski_gravity}.

A special case of a Horndeski gravity is $4\partial ST$ 
(``4 derivative Scalar-Tensor'', alias
``Einstein scalar Gauss-Bonnet'') gravity
\begin{align}
   \label{eq:action_4dST_gravity}
   S_{4\partial ST}
   \equiv
   \int d^4x\sqrt{-g}\left(
      \frac{1}{2}R 
      + 
      X
      -
      V\left(\phi\right)
      +
      \alpha\left(\phi\right)X^2
      +
      \beta\left(\phi\right)\mathcal{R}_{GB}
   \right)
   ,
\end{align}
where 
$\mathcal{R}_{GB}
\equiv 
R^2 
- 
4R_{\mu\nu}R^{\mu\nu} 
+ 
R_{\mu\alpha\nu\beta}R^{\mu\alpha\nu\beta}$
is the Gauss-Bonnet scalar. While the Gauss-Bonnet scalar does
not appear in the action \eqref{eq:general_horndeski_action},
the equations of motion for $4\partial ST$ gravity are second order
in time and 
spatial derivatives\cite{Zwiebach:1985uq,Gross:1986mw,
Kovacs:2020pns,Kovacs:2020ywu}.
In fact one can show through field redefinitions that
the Gauss-Bonnet scalar term can be transformed into the 
form presented in 
Eq.~\eqref{eq:general_horndeski_action}\cite{Kobayashi:2011nu}.
This theory has attracted attention as for some coupling functions
$\beta\left(\phi\right)$, black holes can have 
scalar hair\cite{Kanti:1995vq,
Sotiriou:2013qea,
Sotiriou:2014pfa,
Sotiriou:2015pka,
Silva:2017uqg,
Doneva:2017bvd,
Minamitsuji:2018xde
}.

Generically one may expect the scalar field $\phi$ to 
\emph{conformally couple} to 
matter fields in the following way\cite{fujii_maeda_2003,will_2018} 
\begin{align}
   S_{(M)}
   =
   \int d^4x\sqrt{-g}L_{(M)}\left[
      \Psi_{(M)},A^2\left(\phi\right)g_{\mu\nu}
   \right]
   ,
\end{align}
where $L_{(M)}$ is the ``matter Lagrangian'', $\Psi_{(M)}$ stands for
all matter fields, and $A^2\left(\phi\right)$
is a positive definite function of $\phi$.
Through \emph{Weyl transformations}\footnote{Weyl transformations
are also sometimes called \emph{conformal transformations},
although we avoid that terminology here, as that 
can also refer to conformal coordinate transformations, where the
coordinates themselves $x^{\mu}$ are transformed so that that
$g_{\mu\nu}\to\Omega g_{\mu\nu}$.}, that is field redefinitions
of the form $g_{\mu\nu}\to\Omega\left(\phi\right)\tilde{g}_{\mu\nu}$
one can redefine the metric to remove the factor $A^2$
(by choosing $\Omega=1/A^2$),
at the expense of adding a scalar field functional prefactor
$1/A^2$ in front of the Ricci scalar in $\mathcal{L}_1$,
and having to redefine the other Horndeski functions 
$\mathcal{G}_i$\footnote{We
note that in general a Weyl transformation will transform $4\partial ST$
gravity to a more general Horndeski gravity theory, 
due the transformation properties of the 
Gauss-Bonnet scalar\cite{Maeda:2009uy}.}.
If there no coupling of the scalar field to
the Ricci scalar of the form $f\left(\phi\right)R$, the action
is said to be in the \emph{Einstein frame}\cite{Hawking:1972qk}.
If there is no conformal coupling between the metric and $\phi$
in the matter action, the action is said to be in the
\emph{Jordan frame}.
For more discussion, see \refcite{Flanagan:2004bz}.
This terminology
was invented before the Horndeski theories became better known,
so the Einstein and Jordan frames do not exhaustively describe all
possible couplings such as those found in $\mathcal{L}_4$ and $\mathcal{L}_5$.
For more discussion about the Jordan and Einstein frames
in the context of classical scalar-tensor gravity, see \refcite{fujii_maeda_2003}.
The form of the coupling $A$
can have a dramatic effect on the potential observational effects
of a given theory\cite{Damour:1996ke,fujii_maeda_2003,Will:2014kxa,will_2018}. 

Finally we mention that in fact one can have an even more general
\emph{disformal coupling} between the metric and 
scalar field\cite{Bekenstein:1992pj}, which do not introduce
any new degrees of freedom to the action:
\begin{align}
   S_{(M)}
   =
   \int d^4x\sqrt{-g}L_{(M)}\left[
      \Psi_{(M)},A^2\left(\phi,X\right)
      \left(
         g_{\mu\nu}
         +
         B^2\left(\phi,X\right)\nabla_{\mu}\phi\nabla_{\nu}\phi
      \right)
   \right]
   .
\end{align}
Here $A^2$ and $B^2$ are positive functions of $\phi,X$.
While disformal couplings were proposed several decades 
ago\cite{Bekenstein:1992pj}, they remain relatively less well studied
than conformal couplings.
Moreover, the Horndeski theories are not invariant under \emph{disformal
transformations}, that is field redefinitions of the form 
$g_{\mu\nu}\to
      A^2\left(\phi,X\right)
      \left(
         g_{\mu\nu}
         +
         B^2\left(\phi,X\right)\nabla_{\mu}\phi\nabla_{\nu}\phi
      \right)$.
Under disformal transformations the Horndeski Lagrangian can transform
to have terms that have higher than second order derivatives.
New degrees of freedom are not necessarily introduced by these
higher derivative terms though; 
for more discussion see 
Refs.~\refcite{Zumalacarregui:2013pma,Kobayashi:2019hrl}.
We note that
the general form of the Horndeski gravity theories are preserved under
\emph{special disformal transformations}, that is transformations
where $A,B$ are only functions of $\phi$\cite{Bettoni:2013diz}.
\section{Local well-posedness of the equations of motion
\label{sec:local_well_posedness}
}
\subsection{General considerations\label{sec:general_considerations}}
	
Local well posedness of the equations of motion is necessary for
constructing numerical solutions.
For hyperbolic partial differential equations,
the theory must have a well-posed \emph{initial value problem} (IVP).
For elliptic partial differential equations, the theory must
have a well posed \emph{boundary value problem} (BVP).
A system of partial differential equations for a given problem setup
are well posed if\cite{evans2010partial}
\begin{enumerate}
\item The problem has a solution 
\item The solution is unique
\item The solution depends continuously on the
   given data for the problem (for example, for an initial value problem
      this would be initial data).
\end{enumerate}
   The last criteria is somewhat subtle, as continuity needs to be 
defined with respect to function space.
A theory that has a well-posed IVP or BVP with respect
to one function space may not be well posed in another\footnote{Typically 
well-posedness for PDE in physics are defined with respect to
a \emph{Sobolev space} such as $\mathbb{H}^2$,
which is the space of all
functions that have weak partial derivatives
of at least second order that are square integrable ($L^2$); 
for more discussion see 
\refcite{evans2010partial,Sarbach:2012pr,Hilditch:2013sba}.}.

As we review in \ref{sec:review_hyperbolicity}, the equations of motion
for Horndeski gravity have a well-posed IVP provided they form
a strongly hyperbolic system.
As in GR\cite{christodoulou2008mathematical,Sarbach:2012pr,Reula_review}, 
the hyperbolicity properties of the Horndeski equations
of motion are formulation-dependent and gauge-dependent.
By ``formulation'', we mean the choice of evolution variables 
and how the equations of motion are written
(for example, the BSSN formulation\cite{Shibata:1995we,Baumgarte:1998te} 
and modified harmonic 
formulation\cite{Friedrich:1996hq,Garfinkle:2001ni,Pretorius:2006tp}),
and by ``gauge'' we mean the choice
of the four coordinate (gauge) degrees of freedom
(for example, harmonic gauge\cite{choquet_bruhat_harmonic}).
The choice of a suitable formulation and gauge is called a
\emph{hyperbolic reduction}\cite{Friedrich:1996hq}.

Since Choquet-Bruhat's work on the Einstein equations in harmonic
coordinates\cite{choquet_bruhat_harmonic},
many strongly-hyperbolic reductions of the Einstein equations have
been found (for reviews, see \refcite{Reula_review,Sarbach:2012pr,Hilditch:2013sba}).
Perturbatively solving the Horndeski equations about
GR also leads to well-posed evolution, as the principal part of the
perturbative equations remain the same as for GR to all orders
in perturbation theory. We discuss the perturbative method
in more detail in Sec.~\ref{sec:perturbative_well_posedness}.

Up until recently, it was unclear if it was possible to
formulate a hyperbolic reduction of the full, 
general Horndeski equations of motion.
Given the earlier success of the generalized harmonic
formulation in evolving the Einstein 
equations\cite{Garfinkle:2001ni,Pretorius:2004jg,Pretorius:2005gq,Pretorius:2006tp}
(under which the Einstein equations are strongly hyperbolic\cite{Friedrich:1996hq})
Papallo and Reall investigated the hyperbolicity of the
Horndeski equations in the generalized harmonic formulation.
They found that for all but a small subset of the theories
(in particular, those with $\mathcal{G}_4=\mathcal{G}_5=0$),
the equations of motion were only weakly, but not strongly
hyperbolic\cite{Papallo:2017qvl,Papallo:2017ddx}.
Recently though, Kovacs and Reall have introduced the
\emph{modified generalized harmonic} (MGH) formulation, in which 
Horndeski equations of motion do form a 
strongly hyperbolic system\cite{Kovacs:2020pns,Kovacs:2020ywu}, provided
the Horndeski terms in the solution are \emph{weakly coupled}.
Currently the MGH formulation is the
only known strongly hyperbolic formulation for the 
full, general Horndeski equations of motion.
Moreover, there are other strongly hyperbolic formulations of
specific Horndeski gravity theories;  
see Sec.~\ref{sec:local_well_posedness_exact_eom}.

\subsection{Weakly-coupled regime
\label{sec:weak_coupling}
}
Before continuing to the various hyperbolic reductions of the
Horndeski equations, we first define what is meant
by a weakly coupled solution.
This condition demands that the smallest length scale as defined
by the spacetime curvature is large compared to the length scales defined
by the Horndeski coupling constants.
The weak coupling condition
has also been called the \emph{weak background field} condition
\cite{Papallo:2017qvl}.  
We can make these conditions more concrete by considering
an orthonormal basis $\{e_{\mu}\}$ (we assume we have also chosen a
foliation of spacelike hypersurfaces $\left\{\Sigma_t\right\}$, 
and that the $e_0$ are chosen to be orthogonal to 
each $t=const.$ hypersurface).
We define the function
$M_t\left[T\right]$ as giving the magnitude of of the largest
component of a tensor $T$ with respect to this basis on the
leaf $\Sigma_t$, and define the length scales
\begin{align}
   M_t\left[R^{\alpha}{}_{\mu\beta\nu}\right]
   \equiv
   \frac{1}{L_R^2}
   ,\qquad
   M_t\left[\nabla_{\mu}\phi\right]
   \equiv
   \frac{1}{L_1}
   ,\qquad
   M_t\left[\nabla_{\mu}\nabla_{\nu}\phi\right]
   \equiv&
   \frac{1}{L_2^2}
   .
\end{align}
We then define the shortest length scale as
\begin{align}
   \frac{1}{L}
   \equiv
   \max\left[\frac{1}{L_R},\frac{1}{L_1},\frac{1}{L_2}\right]
   .
\end{align}
The weak coupling condition is obeyed on a leaf $\Sigma_t$ if the following
conditions are obeyed
\begin{subequations}
\label{eq:weak_coupling_conditions}
\begin{align}
   \frac{1}{L^{2k-2}}
   \left|\partial_X^k\mathcal{G}_2\right|
   \ll
   1
   ,&\qquad
   k=1,2
   \\
   \frac{1}{L^{2k}}
   \left|\partial_X^k\partial_{\phi}^l\mathcal{G}_3\right|
   \ll
   1
   ,&\qquad
   k=0,1,2,\;\;\;\qquad l=0,1,\;\;\; \qquad 1\leq k+l\leq 2
   \\
   \frac{1}{L^{2k}}
   \left|\partial_X^k\partial_{\phi}^l\mathcal{G}_4\right|
   \ll
   1
   ,&\qquad
   k=0,1,2,3,\qquad l=0,1,2,\qquad k+l \leq 3
   \\
   \frac{1}{L^{2k+2}}
   \left|\partial_X^k\partial_{\phi}^l\mathcal{G}_5\right|
   \ll
   1
   ,&\qquad
   k=0,1,2,3,\qquad l=0,1,2,\qquad 1\leq k+l \leq 3
   .
\end{align}
\end{subequations}
From the Horndeski equations of motion 
(see ~\ref{sec:eom_horndeski_gravity}), we see these conditions
imply that the Horndeski terms in the equations of motion
are ``small'' compared to the shortest curvature/scalar gradient length scale. 
Solutions that are not weakly-coupled are called \emph{strongly-coupled}.
There is currently
no consensus on how to interpret strongly-coupled solutions
to the Horndeski gravity theories. 
Clearly, some Horndeski gravity theories can have well-defined strongly coupled
classical solutions (for a simple example: minimally coupled scalar fields in GR
can be classically evolved up to the formation of 
curvature singularities\cite{Choptuik:1992jv}).
There are some Horndeski theories that appear to break down though in the
strong coupling regime, such as $4\partial ST$ gravity with a nonzero
Gauss-Bonnet coupling 
$\beta$\cite{Ripley:2019hxt,Ripley:2019irj,Ripley:2019aqj,Ripley:2020vpk,
East:2021bqk,Corelli:2022pio,Corelli:2022phw}.
Whether or not a given Horndeski theory has sensible strongly-coupled solution
will likely need to be determined on a case-by-case basis.
We will largely restrict our discussion to weakly-coupled solutions
in this article, although we will mention numerical work on solutions 
that exhibit the Vainshtein mechanism, which necessarily lies
in the strongly-coupled regime\cite{VAINSHTEIN1972393,Joyce:2014kja}.
Note that black holes (and black hole mergers) can be in the weakly-coupled
regime for a Horndeski gravity theory, so long as the curvature scale
set by the black hole (such as the black hole radius) is large compared
to the scale set by the Horndeski coupling constants.
More generally, Horndeski solutions that describe gravity in the 
strong field, dynamical regime can also be weakly coupled so long
as the Horndeski coupling constants are smaller than the smallest curvature
scale set by the solution.

\section{Local well-posedness of the \emph{perturbative} equations of motion
\label{sec:perturbative_well_posedness}
}
We first consider the Horndeski equations of motion rewritten in
a \emph{naive perturbation theory} (or \emph{order-reduction}) scheme.
The main advantage of the naive-perturbative method is that it reduces
the principal part of the Horndeski equations of motion to that
of GR, so that methods already used in numerical relativity can be
directly applied.
The main disadvantage of the naive-perturbative approach is the presence 
for so-called \emph{secular instabilities}, which 
can spoil the accuracy of higher order perturbative solutions that are integrated
over sufficiently long time scales.

In the perturbative approach,
the metric and scalar field are expanded as a power-series
in a small parameter $\epsilon$ about a GR background
\begin{subequations}
   \label{eq:perturbative_expansion}
   \begin{align}
      g_{\mu\nu}
      &=
      g_{\mu\nu}^{(0)}
      +
      \sum_{k=1}^{\infty}\epsilon^k g^{(k)}_{\mu\nu}
      ,\\
      \phi
      &=
      \phi^{(0)}
      +
      \sum_{k=1}^{\infty}\epsilon^k\phi^{(k)}
      .
   \end{align}
\end{subequations}
Here $\epsilon$ is a dimensionless number; it is used to keep track
of the perturbative order one is working in, and can be set to $\epsilon=1$
before evaluating an expression at a given order.
The expressions \eqref{eq:perturbative_expansion} are inserted
into the full Horndeski equations of motion
$E^{(g)}_{\mu\nu}$ and $E^{(\phi)}$ (see \ref{sec:eom_horndeski_gravity}), 
and truncated at each order in $\epsilon$.
This results in a ``tower'' of partial differential equations, that
schematically take the form
\begin{align}
   \label{eq:perturbatve_tower}
   &\epsilon^0E^{(g,0)}_{\mu\nu}\left[g^{(0)}_{\mu\nu},\phi^{(0)}\right]
   =
   0
   ,
   &\epsilon^0 E^{(\phi,0)}\left[g^{(0)}_{\mu\nu},\phi^{(0)}\right]
   =
   0
   ,\nonumber\\
   &\qquad\qquad\qquad\qquad\qquad\qquad\vdots
   \\
   &\epsilon^kE^{(g,k)}_{\mu\nu}
   \left[
      g^{(0)}_{\mu\nu},\phi^{(0)},
      \cdots,
      g^{(k)}_{\mu\nu},\phi^{(k)}
   \right]
   =
   0
   ,
   &\epsilon^k E^{(\phi,k)}
   \left[
      g^{(0)}_{\mu\nu},\phi^{(0)},
      \cdots,
      g^{(k)}_{\mu\nu},\phi^{(k)}
   \right]
   =
   0
   ,
   \nonumber\\
   &\qquad\qquad\qquad\qquad\qquad\qquad\vdots
\end{align}
Moreover, we assume that each Horndeski term (beyond $\mathcal{L}_1$)
appears at least at order $\epsilon$
\begin{align}
   \label{eq:small_coupling_approx}
   \mathcal{L}_{i\geq2} \sim \epsilon^{p_i}, \qquad p_i\geq1
   .
\end{align}
This is easy to accomplish by
for example setting the couplings constants to be $\sim \epsilon^{p_i}$.
We note that this is a distinct condition from
a solution  to the Horndeski theories being weakly coupled
(see Eq.~\eqref{eq:weak_coupling_conditions}). 
From Eqs.~\eqref{eq:perturbative_expansion},~\eqref{eq:small_coupling_approx},
the principal part of the
perturbative equations of motion \eqref{eq:perturbatve_tower}, 
to every order in $k$, is the same as the principal part of GR 
(the $\mathcal{L}_1$ term).
This can been seen by consider the full Horndeski equations of motion,
which given Eq.~\eqref{eq:small_coupling_approx} take the form
\begin{align}
   \left(E^{(g)}\right)^{\alpha}_{\beta}
   &=
   -
   \frac{1}{4}\delta^{\alpha\gamma_1\gamma_2}_{\beta\delta_1\delta_2}
   R_{\gamma_1\gamma_2}{}^{\delta_1\delta_2}
   -
   \left(V+X\right)\delta^{\alpha}_{\beta}
   -
   \nabla^{\alpha}\phi\nabla_{\beta}\phi
   +
   \mathcal{O}\left(\epsilon^p\right)
   \nonumber\\
   &\equiv
   \left(E^{(g)}_{GR}\right)^{\alpha}_{\beta}
   +
   \mathcal{O}\left(\epsilon^p\right)
   ,\\
   E^{(\phi)}
   &=
   -
   \Box\phi
   +
   \partial_{\phi}V
   +
   \mathcal{O}\left(\epsilon^p\right)
   \nonumber\\
   &\equiv
   E^{(\phi}_{GR}
   +
   \mathcal{O}\left(\epsilon^p\right)
   .
\end{align}
To each order in $k$, the principal part for the evolved variables
$g^{(k)}_{\mu\nu},\phi^{(k)}$ is the same as it is for GR.
Plugging in \eqref{eq:perturbative_expansion}, to each order in $k$
we have
\begin{align}
   \left(
      E^{(g)}_{GR}
      \left[g^{(k)}_{\mu\nu},\phi^{(k)}\right]
   \right)^{\alpha}_{\beta}
   +
   \left(F^{(k)}\left[
      g^{(0)}_{\mu\nu},\phi^{(0)},
      \cdots,
      g^{(k-1)}_{\mu\nu},\phi^{(k-1)}
   \right]\right)^{\alpha}_{\beta}
   &=
   0
   ,\\
   E^{(\phi}_{GR}
   \left[g^{(k)}_{\mu\nu},\phi^{(k)}\right]
   +
   F^{(\phi)}\left[
      g^{(0)}_{\mu\nu},\phi^{(0)},
      \cdots,
      g^{(k-1)}_{\mu\nu},\phi^{(k-1)}
   \right]
   &=
   0
   ,
\end{align}
where $\left(F^{(g,k)}\right)^{\alpha}_{\beta}$ and
$F^{(\phi,k)}$ contain only up to first derivatives of their arguments.
The perturbative equations of motion to each order in $k$
can then be evolved in a well-posed manner using the same formulations
of the equations of motion used in numerical evolutions of
the Einstein equations.

A common assumption is to set $\phi^{(0)}=0$, and to solve only the first
order corrections to the equations of motion.
From the Horndeski equations of motion and using the
conditions \eqref{eq:coefficient_assumptions}, 
we see that with $\phi^{(0)}=0$
then there are no corrections to the linearly corrected
Einstein (tensor) equations $E^{(g,1)}_{\mu\nu}$
(assuming $V(0)=0$), in which case we can set 
$g_{\mu\nu}^{(1)}=0$.
The linearly corrected
scalar equations $E^{(\phi,1)}$
then only contain $\phi^{(1)}$ (and not $g_{\mu\nu}^{(1)}$), so we
can solve for the scalar field on a GR background.
This approach is often called the \emph{decoupling approximation}.
The scalar field back-reaction on the metric appears at second order
in the perturbative expansion\footnote{The correction to the
metric \emph{will} generally 
occur at least at second order in the perturbative expansion,
due to the presence of $X$ in $\mathcal{L}_1$.}.

The naive-perturbative approach has been applied to some Horndeski 
gravity theories\cite{Witek:2018dmd,Okounkova:2019zep,Okounkova:2020rqw}.
One of the most attractive features though of the formalism is that
it can be applied to essentially any modified theory of gravity,
including theories whose exact equations of motion may not 
have well-posed initial value formulations, such
as \emph{dynamical Chern-Simons gravity}\cite{Alexander:2009tp,
Yunes:2013dva,Delsate:2014hba,
Okounkova:2017yby,Okounkova:2018abo,Okounkova:2018pql} 
and other higher derivative modified gravity 
theories\cite{Endlich:2017tqa,Cayuso:2020lca}.
As discussed in Sec.~\ref{sec:perturbative_secular_instabilities}, 
the presence of secularly growing
error terms can complicate the interpretation of higher order
perturbative solutions.
In that section 
we discuss how the recently proposed
\emph{dynamical renormalization group} could potentially address
the problem of secular growth in perturbative numerical solutions.

\section{Local well-posedness of the \emph{exact} equations of motion
\label{sec:local_well_posedness_exact_eom}
}
We next consider well-posed formulations of the exact
equations of motion for the Horndeski theories of gravity; that is
for the complete set of equations listed in \ref{sec:eom_horndeski_gravity}.
While exact solutions do not suffer from secular instabilities, 
one has to contend with finding a well-posed formulation of
the equations of motion. We review well-posed formulations
for various special cases of the Horndeski gravity theories, then
end with a formulation that leads to well-posed evolution
for all of the Horndeski gravity theories: the \emph{modified
generalized harmonic} (MGH) formulation\cite{Kovacs:2020pns,Kovacs:2020ywu}.
\subsection{Well posed formulations of quadratic Horndeski gravity
\label{sec:quadratic_horndeksi_gravity_well_posedness}
}
Quadratic Horndeski gravity
(also often called ``K-essence''
theory\cite{Armendariz-Picon:2000nqq,Armendariz-Picon:2000ulo}),
has the action
\begin{align}
   S=
   \int d^4x\sqrt{-g}\left(
      \mathcal{L}_1
      +
      \mathcal{L}_2
   \right)
   .
\end{align}
The equations of motion for this theory are given in
\ref{sec:eom_horndeski_gravity}, (setting $\mathcal{G}_i=0,i>2$):
\begin{align}
   R_{\mu\nu}
   -
   \frac{1}{2}g_{\mu\nu}R
   +
   \frac{1}{2}\partial_X\mathcal{G}_2
   \nabla_{\mu}\phi\nabla_{\nu}\phi
   -
   \frac{1}{2}g_{\mu\nu}\mathcal{G}_2
   &=
   0
   ,\\
   \nabla_{\mu}\left(\partial_X\mathcal{G}_2\nabla^{\mu}\phi\right)
   +
   \partial_{\phi}\mathcal{G}_2
   &=
   0
   .
\end{align}
For this theory,
the principal part of the tensor equation remains unchanged
from that of GR, and it is straightforward to see that the
scalar equations of motion remain strongly hyperbolic
in the weak-coupling limit (see Ref.~\refcite{Rendall:2005fv}
for a more general discussion). Thus, the standard formulations used
in numerical relativity can be used to evolve this class
of Horndeski gravity theories (if shocks
form in the equations of motion, high resolution shock
capturing methods can be used\cite{Bezares:2021dma}, see
Sec.~\ref{sec:shock_formation}).
\subsection{Well-posed formulation of Cubic Horndeski gravity
\label{sec:cubic_horndeksi_gravity_well_posedness}
}
We next consider cubic Horndeski gravity, which has the action
\begin{align}
   S=
   \int d^4x\sqrt{-g}\left(
      \mathcal{L}_1
      +
      \mathcal{L}_2
      +
      \mathcal{L}_3
   \right)
   .
\end{align}
The equations of motion for this theory are 
given in \ref{sec:eom_horndeski_gravity}
(setting $\mathcal{G}_i=0,i>3$).
Kovacs\cite{Kovacs:2019jqj} pointed out that from the perspective of
local well-posedness in the weakly-coupled limit, 
the terms that could cause problems compared to GR are
$R_{\alpha\beta}\nabla^{\alpha}\nabla^{\beta}\phi$ in the scalar
equations of motion, and terms involving second derivatives of the
scalar field in the tensor equations of motion.
The Ricci tensor can be replaced with second derivatives
of the scalar field by considering the trace-reversed tensor
equations of motion
\begin{align}
   E_{\alpha\beta}^{(g)}
   -
   \frac{1}{2}g_{\alpha\beta}g^{\mu\nu}E_{\mu\nu}^{(g)}
   &=
   R_{\mu\nu}
   \nonumber\\&+
   \frac{1}{2}\left(
      \mathcal{G}_2
      -
      X\partial_X\mathcal{G}_2
      -
      X\partial_X\mathcal{G}_3\Box\phi
   \right)
   g_{\alpha\beta}
   \nonumber\\&-
   \frac{1}{2}\left(
      1
      +
      \partial_X\mathcal{G}_2
      +
      2\partial_{\phi}\mathcal{G}_3
   \right)
   \nabla_{\alpha}\phi\nabla_{\beta}\phi
   \nonumber\\&+
   \frac{1}{2}\partial_X\mathcal{G}_3\left(
      -
      \Box\phi\nabla_{\alpha}\phi\nabla_{\beta}\phi
      +
      2\nabla_{(\alpha}\phi\nabla_{\beta)}\nabla_{\gamma}\phi\nabla^{\gamma}\phi
   \right)
   .
\end{align}
Using this to replace $R_{\mu\nu}$ in the scalar equation of motion,
the principal symbol for the tensor and scalar field takes an upper
triangular structure
(see Kovacs\cite{Kovacs:2019jqj} for the full equations of motion).
With this simplified structure, 
Kovacs was able to prove the local well-posedness of
cubic Horndeski gravity for three popular formulations used in
numerical relativity, 
including the BSSN formulation\cite{Shibata:1995we,Baumgarte:1998te}
and the CCZ4 formulation\cite{Bona:2003fj,Alic:2011gg} 
of the equations of motion.
As we review later, one of these formulations has been used
to evolve (binary) black holes in this 
theory\cite{Figueras:2020dzx,Figueras:2021abd}.
\subsection{Well-posed formulations of
Bergmann--Wagoner scalar-tensor gravity theories 
\label{sec:scalar_tensor_special_case}
}
For completeness, we also mention the original
\emph{scalar-tensor} gravity 
theories\cite{Wagoner:1970vr,nordtvedt1970post,will_2018,Quiros:2019ktw}
\begin{align}
   \label{eq:scalar_tensor_gravity}
   S
   =
   \int d^4x\sqrt{-g}\left(
      F\left(\phi\right) R
      +
      Z\left(\phi\right)X
      -
      V\left(\phi\right)
   \right)
   .
\end{align}
When $F=\phi,Z=\omega/\phi$ (where $\omega$ is a constant),
and $V=0$, this action reduces to
Brans-Dicke gravity\cite{Brans:1961sx}
(in this case the theory limits to GR when $\phi\to1$, $\omega\to\infty$).
Scalar-tensor gravity is a special case of the
$\mathcal{L}_1$, $\mathcal{L}_2$, 
and $\mathcal{L}_4$ Horndeski gravity terms:
$\mathcal{L}_2=\left(Z\left(\phi\right) -1\right)X$,
$\mathcal{L}_4=F\left(\phi\right)-1$,
This theory was shown to have a well-posed initial value formulation
in a BSSN-styled formulation of the theory\cite{Salgado:2008xh}
(although note that the authors in that work defined their
scalar field differently from what we have presented here).
Provided there is no coupling of the form $A^2\left(\phi\right)g_{\mu\nu}$
in the matter Lagrangian, the action \eqref{eq:scalar_tensor_gravity}
is said to be in the \emph{Jordan frame}.
One can generally perform a Weyl transformation
(a field redefinition of the form
$g_{\mu\nu}\to \Omega\left(\phi\right)\tilde{g}_{\mu\nu}$) to set
$F\left(\phi\right)R\to \tilde{R}$; that is to go to the
\emph{Einstein frame} (see for example Ref.~\refcite{fujii_maeda_2003}). 
In this frame, there is no mixing
between the scalar and tensor degrees of freedom in the
principal symbol.
The well posedness of the equations of motion in the Einstein
frame is then straightforward
to show, and the theory can essentially be 
solved numerically using the same methods that are used
to solve for the GR with a minimally coupled
scalar field with a potential. 
Because of this, most researchers
in numerical relativity evolve black holes and neutron
stars in the Einstein frame; recent work includes 
\refcite{terHaar:2020xxb,Bezares:2021yek,Bezares:2021dma}.

\subsection{A well-posed formulation for all Horndeski gravity
theories: the modified generalized harmonic (MGH) formulation
\label{sec:mgh_formulation}
}

There is currently only one known strongly hyperbolic
formulation for general weakly-coupled Horndeski gravity
theories ($\mathcal{L}_1$ through $\mathcal{L}_5$): the 
\emph{modified generalized harmonic (MGH) 
formulation}\cite{Kovacs:2020pns,Kovacs:2020ywu}\footnote{\emph{Note added
after publication}: the MGH formulation has been adapted
to the CCZ4 formulation by
Arest\'{e}, Clough, and Figueras\cite{AresteSalo:2022hua}.
In that work the authors also evolved binary black holes through
merger in $4\partial$ST gravity, using singularity-avoiding coordinates.}.
The setup for the formulation is:
in a Lorentzian spacetime $(M,g)$, we introduce two auxiliary
Lorentzian metrics $\tilde{g}^{\alpha\beta}$ and 
$\hat{g}^{\alpha\beta}$.
We will always raise and lower indices with
the spacetime metric $g_{\alpha\beta}$,
so for example $\hat{g}^{\alpha\beta}\equiv
g^{\alpha\gamma}g^{\beta\delta}\hat{g}_{\gamma\delta}$.
We also define the traces 
$\tilde{g}\equiv \tilde{g}^{\alpha\beta}g_{\alpha\beta}$ and
$\hat{g}\equiv\hat{g}^{\alpha\beta}g_{\alpha\beta}$.
The MGH formulation imposes the following
conditions on the coordinate functions $x^{\gamma}$:
\begin{align}
\label{eq:mh_condition}
   C^{\gamma}
   \equiv&
   H^{\gamma}
   -	
   \tilde{g}^{\alpha\beta}
   \nabla_{\alpha}\nabla_{\beta}x^{\gamma}
   \nonumber\\
   =&
   H^{\gamma}
   +	
   \tilde{g}^{\alpha\beta}\Gamma_{\alpha\beta}^{\gamma}
   \dot{=}
   0
   .
\end{align} 
We put a ``dot'' over the last equals sign to indicate that numerically,
one generally expects $C^{\gamma}$ to not exactly equal zero
due to truncation error.
As in the generalized harmonic 
formulation\cite{Friedrich:1996hq,Garfinkle:2001ni,Pretorius:2004jg},
$H^{\gamma}$ are the source functions that, 
along with $\tilde{g}^{\alpha\beta}$ and $\hat{g}^{\alpha\beta}$,
specify the gauge, and $C^{\gamma}$ 
is called the \emph{constraint violation}.
We next define the MGH equations of motion as 
\begin{align}
\label{eq:basic_mgh_eom}
   \left(E^{(g,MGH)}\right)^{\alpha\beta}
   \equiv
   &
   \left(E^{(g)}\right)^{\alpha\beta}
   -	
   \hat{P}_{\delta}{}^{\gamma\alpha\beta}
   \nabla_{\gamma}C^{\delta}
\nonumber\\
   &
   - 
   \frac{1}{2}\kappa\left(
      n^{\alpha}C^{\beta}
      +	
      n^{\beta}C^{\alpha}
      +	
      \rho n^{\gamma}C_{\gamma} g^{\alpha\beta}
   \right)
   =
   0
   ,
\end{align}
where $n^{\alpha}$ is a time-like vector 
(we assume $n^{\alpha}$ is timelike with respect to $g^{\alpha\beta}$,
$\tilde{g}^{\alpha\beta}$, and 
$\hat{g}^{\alpha\beta}$), and
\begin{align}
   \hat{P}_{\delta}{}^{\gamma\alpha\beta}
   \equiv
   \frac{1}{2}\left(
      \delta_{\delta}^{\alpha}\hat{g}^{\beta\gamma}
      +	
      \delta_{\delta}^{\beta}\hat{g}^{\alpha\gamma}
      -	
      \delta_{\delta}^{\gamma}\hat{g}^{\alpha\beta}
   \right)
   .
\end{align}
In Eq.~\eqref{eq:basic_mgh_eom}
we included constraint damping terms with the constants 
$\kappa$ and $\rho$; see \refcite{Gundlach:2005eh}. 
These terms would not affect the solution if $C^{\gamma}$ was
exactly zero,
but in numerical applications this is generally not the case, 
and without damping terms
the constraint violation could blow up exponentially in 
time\cite{Pretorius:2004jg,Pretorius:2005gq,Pretorius:2006tp,East:2020hgw}.
\footnote{Note that Eq.~\eqref{eq:basic_mgh_eom} is slightly different
from Kovacs and Reall\cite{Kovacs:2020ywu}. Here we use 
$\nabla_{\gamma}C^{\delta}$ instead of
$\partial_{\gamma}C^{\delta}$.
We choose this form so that the MGH equations we introduce
here matches that of
the standard generalized harmonic formulation\cite{East:2020hgw}
in the limit 
$\hat{g}^{\alpha\beta}=\tilde{g}^{\alpha\beta}=g^{\alpha\beta}$.
}.
From Eq.~\eqref{eq:mh_condition},
we see that in the MGH formulation the coordinates
$x^{\alpha}$ obey a hyperbolic equation with characteristics determined by
$\tilde{g}^{\alpha\beta}$. Taking the divergence of 
Eq.~\eqref{eq:basic_mgh_eom},
and assuming 
$\nabla_{\alpha}E^{\alpha\beta}=0$ (which by general covariance 
of the action, holds for the Horndeski equations of motion),
we obtain a hyperbolic equation
for the constraint violating modes $C^{\alpha}$: 
\begin{align}
\label{eq:eom_constraint_violating}
   -  
   \frac{1}{2}\hat{g}^{\alpha\gamma}
   \nabla_{\alpha}\nabla_{\gamma}C^{\beta}
   -  
   \hat{g}^{\gamma\beta}R_{\delta\gamma}C^{\delta}
   -  
   \left(
      \nabla_{\alpha}\hat{P}_{\delta}{}^{\gamma\alpha\beta}
   \right)
   \left(
      \nabla_{\gamma}C^{\delta}
   \right)
   \nonumber \\
   -  
   \frac{1}{2}\kappa\nabla_{\alpha}\left(
      n^{\alpha}C^{\beta}
      +  
      n^{\beta}C^{\alpha}
      +  
      \rho n^{\gamma}C_{\gamma} g^{\alpha\beta}
   \right)
   =
   0
   .
\end{align}
From Eq.~\eqref{eq:eom_constraint_violating},
we see that the constraint violating modes 
obey a hyperbolic equation with characteristics determined
by $\hat{g}^{\alpha\beta}$.

Kovacs and Reall proved that in the weak coupling regime that the
Horndeski equations of motion in the MGH formulation:
\begin{align}
   \left(E^{(g,MGH)}\right)^{\mu\nu}
   = 0
   ,\qquad
   E^{(\phi)}
   =
   0
   ,
\end{align}
form a strongly hyperbolic system of 
partial differential equations\cite{Kovacs:2020pns,Kovacs:2020ywu}.
Specifying a gauge in the MGH formulation amounts to choosing the
functional form of  
$\tilde{g}^{\alpha\beta}$, 
$\hat{g}^{\alpha\beta}$, and $H^{\gamma}$.
With the choice
$\tilde{g}^{\alpha\beta}
=\hat{g}^{\alpha\beta}
=g^{\alpha\beta}$, the MGH 
formulation reduces to the generalized harmonic formulation.

We next outline the main elements of Kovacs and Reall's
strong hyperbolicity argument. 
In the GH formulation, 
the coordinate degrees of freedom and the constraint violation
degrees of freedom all have the same characteristics:
they are determined by $g^{\mu\nu}$.
The equations of motion for GR with minimally coupled matter fields 
in the GH formulation are remarkable 
as despite the degeneracy of the eigenvalues of the principal symbol,
the principal symbol is diagonalizable.
This is not the case of the Horndeski gravity theories: as
Papallo and Reall showed\cite{Papallo:2017qvl,Papallo:2017ddx},
there remains a Jordan block in the principal symbol of the
equations of motion for these theories, that is locally
the principal symbol takes the form
\begin{align}
   \mathcal{P}^{(GH)}\left(\xi\right)
   =
   \begin{pmatrix}
      c_1 & 0      & 0      & \cdots   \\
      0   & c_1    & 0      & \cdots   \\
      0   & 0      & \ddots            \\
          & \cdots & 0      & c_n & 1  \\
          & \cdots & 0      & 0   & c_n
   \end{pmatrix}
   ,
\end{align}
where the $c_i$ are the characteristic speeds. 
Papallo and Reall found this Jordan block cannot
be removed for a single constraint violating mode.
One of the key insights of Kovacs and Reall was to notice that the
coordinates and constraint violation fields did not
need to propagate on the spacetime metric, 
as they are unphysical.
The introduction of the auxiliary metrics $\tilde{g}^{\mu\nu}$
and $\hat{g}^{\mu\nu}$ breaks some of the degeneracies in the
characteristic speeds in the principal symbol
among the coordinate and gauge-violating degrees of freedom, 
which they showed is sufficient to make the principal symbol diagonalizable, 
at least for weakly-coupled solutions\cite{Kovacs:2020pns,Kovacs:2020ywu}.
To gain some intuition as to how breaking degeneracies in the eigenvalues
of the principal symbol could help, recall that perturbing
the eigenvalues of a Jordan-form matrix generally makes it
diagonalizable, for example 
\begin{align}
   \begin{pmatrix}
      c_n & 1 \\
      0   & c_n+\epsilon
   \end{pmatrix}
\end{align}
is diagonalizable 
with eigenvectors $(1,0)$ and $(1,\epsilon)$.

The work of Kovacs and Reall suggests that it should be possible
to generalize other formulations used to evolve the Einstein
equations, for example the BSSN formulation of the equations of 
motion\cite{Shibata:1995we,Baumgarte:1998te},
to give a strongly hyperbolic formulation for the
Horndeski equations of motion for weakly-coupled solutions.

\section{Exact solutions to the \emph{fixed}
equations of motion\label{sec:fixing_eqns}}

A related method to solving the full Horndeski equations is to
modify (\emph{fix}) the Horndeski equations in such as way that may allow
for a well-posed initial value problem, even outside the 
weak-coupling regime.
The \emph{equation-fixing} method takes inspiration from the so-called
Israel-Stewart method\cite{Israel:1976tn,1976PhLA...58..213I,Israel:1979wp} 
to render the relativistic Navier-Stokes
equations into a strongly-hyperbolic system of 
equations\cite{Cayuso:2017iqc,Allwright:2018rut}.
Fixing the equations involves modifying them in such a
way so that only high-energy degrees of freedom are modified, 
and the dynamics of
low-energy degrees of freedom remain essentially unchanged.
If we view the equations of motion of Horndeski theory as describing
a low-energy effective deviation from the Einstein equations,
then the essential physics of the theory should be captured
by the low-energy degrees of freedom anyways.
Moreover, hyperbolicity/ellipticity is defined with respect
to the principal part, which is most sensitive to high-frequency
(high-energy) degrees of freedom, so changing that part of the
equations of motion may render them strongly hyperbolic.

Fixing the equations may allow for a strongly
hyperbolic reduction even for strongly coupled solutions
(although, as mentioned earlier, this
regime may lie outside the regime of applicability of
the theory).
To illustrate the idea of fixing in this situation, 
consider the scalar equations
of motion for quadratic Horndeski gravity
\begin{align}
   \label{eq:full_g2_eom}
   \partial_{0}\left(
      \sqrt{-g}\partial_X\mathcal{G}_2\partial^0\phi
   \right)
   +
   \partial_{i}\left(
      \sqrt{-g}\partial_X\mathcal{G}_2\partial^i\phi
   \right)
   +
   \partial_{\phi}\mathcal{G}_2
   =
   0
   .
\end{align}
These equations can break down outside of the weak-coupled
regime\cite{Babichev:2016hys,Babichev:2017lrx}; 
one proposed way to prevent this breakdown
is to fix the equations by introducing a new field $\Pi$
such that the equations of motion are\cite{Bezares:2021yek}
\begin{subequations}
\label{eq:fixed_g2_eom}
\begin{align}
   \partial_{0}\left(
      \sqrt{-g}\Pi \partial^0\phi
   \right)
   +
   \partial_{i}\left(
      \sqrt{-g}\Pi \partial^i\phi
   \right)
   +
   \partial_{\phi}\mathcal{G}_2
   &=
   0
   ,\\
   \partial_0\Pi
   +
   \frac{1}{\tau}\left(\Pi - \partial_X\mathcal{G}_2\right)
   &=
   0
   ,
\end{align}
\end{subequations}
where $\tau>0$ is a new constant, which can be thought of
as a ``relaxation timescale''.
Provided $\tau$ is shorter than the physical timescales that
are being simulated, the ``low-energy mode'' of the solutions 
to the fixed system \eqref{eq:fixed_g2_eom} should
be similar to those of the full system \eqref{eq:full_g2_eom}.
This fix has been found to be helpful for evolving outside
of the weakly-coupled regime in the 
numerical evolution of this class of 
theories\cite{Bezares:2021yek,Bezares:2021dma,Lara:2021piy}.

Beyond quadratic Horndeski gravity, this method has been applied to
cubic Horndeski\cite{Gerhardinger:2022bcw}, and
ESGB gravity\cite{Franchini:2022ukz}.
The fixing approach has also be applied
to higher-derivative theories of gravity\cite{Allwright:2018rut,Cayuso:2020lca}.
Much work remains in systematizing this approach, and quantitatively
determining how much a given ``fix'' affects low-energy degrees
of freedom in a given physical scenario (such as during black hole merger).
There has not yet been any formal investigation of the
well-posedness of the fixed Horndeski evolution and constraint equations.
While it is reasonable to assume that the breakdown in hyperbolicity
of a Horndeski theory indicates that the theory is no longer predictive
in that regime 
(at least without including higher order corrections or a complete theory of say
quantum gravity), there is some debate on how to interpret 
the breakdown\cite{Corelli:2022pio,Corelli:2022phw}.

\section{Challenges to constructing global solutions to
   Horndeski theories
   \label{sec:failure_global}
}
\subsection{Local versus global solutions\label{sec:local_vs_global_solutions}}
As we discussed in
Sec.~\ref{sec:local_well_posedness} and 
Sec.~\ref{sec:local_well_posedness_exact_eom},
for weakly-coupled solutions to the equations of motion, 
the structure of the equations of motion for the Horndeski 
are similar to the structure of the Einstein equations,
and there exists several strongly hyperbolic
formulations of the Horndeski equations of motion which have already
been used in numerical relativity simulations of binary black hole
spacetimes\cite{East:2020hgw,Figueras:2021abd,East:2021bqk}.
While the Horndeski theories have a well-posed initial 
value problem in this regime, there is no guarantee that 
weakly coupled initial data 
will remain weakly coupled at later times.

We first recall that solutions to GR can break down at 
\emph{spacetime singularities}\cite{hawking_ellis_1973}.
\emph{Naked singularities} are not hidden by an event horizon.
Naked singularities can be formed in Einstein gravity
with a minimally coupled scalar
field\cite{10.2307/2118619}
(and even in vacuum GR in four and higher\cite{Zhang:2015rsa} dimensions)
starting from regular, low-curvature initial data.
This being said, 
these solutions are ``unstable''\cite{christodoulou1999instability},
in the sense that small perturbations to the initial data lead to solutions
where there are no naked singularities.
More generally, the 
\emph{weak cosmic censorship conjecture}\cite{1969NCimR...1..252P}
posits that for generic initial data,
solutions to GR coupled to ``standard'' matter fields (for example 
minimally coupled scalar field matter) have well defined global
solutions (e.g. the spacetime is geodesically complete)
outside of trapped regions.
While there remains no proof of this conjecture, no counterexample to it
has ever been found, and it is widely expected to be true\cite{Wald:1997wa}.
A well-defined solution can mean there are no curvature singularities outside
of black hole horizons, 
but more broadly it can be taken to
mean that the Einstein equations can be evolved in time,
for all time and space, without having to introduce new (ad-hoc)
boundary conditions.

Depending on the theory being considered,
exact solutions to a Horndeski theory can suffer
from several different kinds of issues.
\begin{enumerate}
   \item For some Horndeski gravity theories,
      the equations of motion can lose their hyperbolic character
      in regions of high curvature--that is
      \emph{elliptic regions}\cite{Ripley:2019hxt,
         Ripley:2019irj,Figueras:2020dzx} can form.
   
      \item There is some evidence that
      curvature singularities can form outside of trapped region for
      some Horndeski theories\cite{Kanti:1995vq,Sotiriou:2014pfa,
      Silva:2017uqg,Doneva:2017bvd,Antoniou:2017hxj,Kleihaus:2015aje}.
   
   \item Some Horndeski gravity theories have solutions that
      form infinitely steep gradients, that is
      \emph{shocks} or 
      \emph{caustics}\cite{Babichev:2016hys,
         deRham:2016ged, 
         Tanahashi:2017kgn,
         Pasmatsiou:2017vcw,
         Babichev:2017lrx,
         Lara:2021piy
      } can form, even from smooth initial data. 
      We note that shocks can form in perfect-fluid solutions
      to the Einstein equations, so strictly speaking this phenomenon
      is not entirely novel in the context of relativistic physics.
      Moreover, provided the equations of motion can be formulated
      as a conservative set of PDE, shock formation does not necessarily 
      signal the breakdown of the theory.

\end{enumerate}
Only strongly-coupled solutions to some Horndeski theories
have been found to have these features,
although there has been no exhaustive search of all solutions
to all Horndeski theories.
There is some numerical evidence that all of the above listed
phenomena can form from initially weakly-coupled initial data.
This being said,
researchers have numerically evolved binary black hole spacetimes
in quadratic Horndeski gravity, cubic Horndeski gravity,
and in ESGB gravity (the last two theories
which can form naked elliptic regions in 
gravitational collapse\cite{Ripley:2019hxt,
Ripley:2019irj,Bernard:2019fjb,Figueras:2020dzx,Lara:2021piy}) 
through merger, and have found sets of initial data that did not lead to
the formation of a naked elliptic regions or 
shocks\cite{East:2020hgw,East:2021bqk,Figueras:2021abd}.

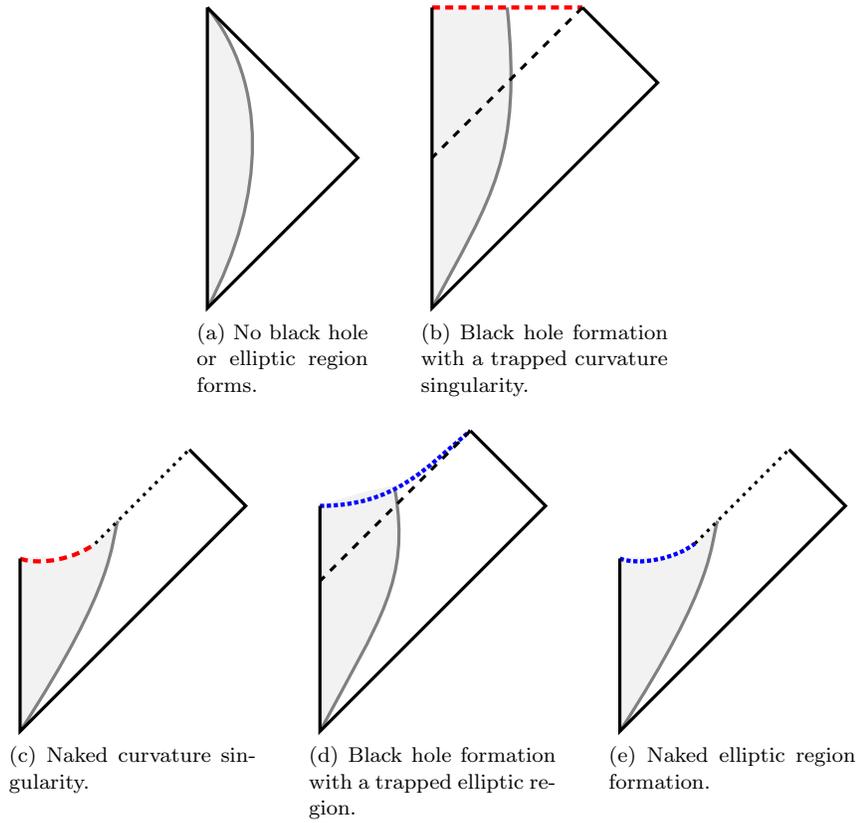
\begin{figure}
   \centering 
   \subfloat[No black hole or elliptic region forms.
      \label{fig:no_black_hole_forms}]{
         \begin{tikzpicture}
            \draw [gray,very thick] 
               (0,0) .. controls (0.8,1.5) and (0.8,3) .. (0,4);
            \fill [gray,opacity=0.1] 
               (0,0) .. controls (0.8,1.5) and (0.8,3) .. (0,4);
            
            \draw [very thick] 
               (0,4) -- (0,0) -- (2,2) -- (0,4);
         \end{tikzpicture}
      }
   \qquad 
   \subfloat[Black hole formation with a trapped curvature singularity.
      \label{fig:standard_black_hole_formation}]{
         \begin{tikzpicture}
            \draw [gray,very thick] 
               (0,0) .. controls (0.8,1.5) and (1.2,2) .. (1,4);
            \fill [gray,opacity=0.1] 
               (0,0) .. controls (0.8,1.5) and (1.2,2) .. (1,4)
               --
               (1,4) -- (0,4);
            
            \draw [red,ultra thick,densely dashed] 
               (0,4) -- (2,4);
            
            \draw [very thick] 
               (0,4) -- (0,0) -- (3,3) -- (2,4);
            \draw [dashed,very thick]
               (2,4) -- (0,2);   
         \end{tikzpicture}
      }
   \\ 
   \subfloat[Naked curvature singularity.
      \label{fig:naked_curvature_singularity}]{
         \begin{tikzpicture}
            \draw [gray,very thick] 
               (0,0) .. controls (1.3,2) and (1.2,2.5) .. (1.3,2.8);
            \fill [gray,opacity=0.1] 
               (0,0) .. controls (1.3,2) and (1.2,2.5) .. (1.3,2.8)
               -- (1,2.5) -- 
               (1,2.5) .. controls (0.25,2.1) and (0.75,2.4) .. (0,2.3);
            
            \draw [red,ultra thick,densely dashed] 
               (0,2.3) .. controls (0.25,2.2) and (0.75,2.3) .. (1,2.5);
            
            \draw [very thick,dotted] 
               (1,2.5) -- (2.25,3.75);
            
            \draw [very thick] 
               (0,2.3) -- (0,0) -- (3,3) -- (2.25,3.75);
         \end{tikzpicture}
      }
   \qquad
   \subfloat[Black hole formation with a trapped elliptic region.
      \label{fig:black_hole_formation}]{
         \begin{tikzpicture}
            \draw [gray,very thick] 
               (0,0) .. controls (0.8,1.5) and (1.2,2) .. (1,3.2);
            \fill [gray,opacity=0.1] 
               (0,0) .. controls (0.8,1.5) and (1.2,2) .. (1,3.2)
               --
               (2,4) .. controls (1,3) and (1.4,3.5) ..  (0,3);
            
            \draw [blue,ultra thick,densely dotted] 
               (0,3) .. controls (1,3) and (1.4,3.5) .. (2,4);
            
            \draw [very thick] 
               (0,3) -- (0,0) -- (3,3) -- (2,4);
            \draw [dashed,very thick]
               (2,4) -- (0,2);   
         \end{tikzpicture}
      }
   \qquad
   \subfloat[Naked elliptic region formation.
      \label{fig:naked_elliptic_formation}]{
         \begin{tikzpicture}
            \draw [gray,very thick] 
               (0,0) .. controls (1.3,2) and (1.2,2.5) .. (1.3,2.8);
            \fill [gray,opacity=0.1] 
               (0,0) .. controls (1.3,2) and (1.2,2.5) .. (1.3,2.8)
               -- (1,2.5) -- 
               (1,2.5) .. controls (0.25,2.1) and (0.75,2.4) .. (0,2.3);
            
            \draw [blue,ultra thick,densely dotted] 
               (0,2.3) .. controls (0.25,2.2) and (0.75,2.3) .. (1,2.5);
            
            \draw [very thick,dotted] 
               (1,2.5) -- (2.25,3.75);
            
            \draw [very thick] 
               (0,2.3) -- (0,0) -- (3,3) -- (2.25,3.75);
         \end{tikzpicture}
      }
   \caption{Schematic Penrose diagram of the different
   kinds of spherical 
   gravitational collapse that have been observed in Horndeski theories,
   starting from weak-field initial data with no initial black hole.
   The gray shaded region depicts the region where there is
   a nonzero scalar field value $\phi$,
   that is we depict gravitational collapse that is driven
   by the Horndeski scalar field.
   A black dashed line indicates the location of an event horizon,
   and the black dotted line indicates a region that would need
   to be excised from the numerical domain in order to main
   well-posed evolution in an untrapped region.
   A red dashed lines indicates a curvature singularity, while a
   blue dotted line indicates a sonic line.
   Panels \eqref{fig:no_black_hole_forms} and 
   \eqref{fig:standard_black_hole_formation} describe
   collapse that has been observed in GR coupled to ``standard''
   matter fields and some Horndeski gravity theories.
   Other Horndeski gravity theories, such as 
   variants of ESGB gravity and cubic Horndeski
   gravity theory, can form (naked) curvature
   singularities and (naked) elliptic 
   regions\cite{Ripley:2019hxt,Ripley:2019irj,Ripley:2019aqj,Ripley:2020vpk,
   Figueras:2020dzx,Figueras:2021abd,East:2021bqk,Corelli:2022phw,Corelli:2022pio}.
   The formation of either signals the solution has exited the
   weakly-coupled regime.
   It is worth noting that clearly not all Horndeski theories will
   form elliptic regions, for example GR with minimally coupled
   scalar field is a Horndeski gravity theory.
   \\
   \\
   \\
   \\
   \\
   }
   \label{fig:penrose_diagram_bh}.
\end{figure}

Numerical studies of the detailed PDE nonlinear
properties of the Horndeski
gravity theories have mostly been confined to spherically
symmetric spacetimes.
In Fig.~\ref{fig:penrose_diagram_bh} we show 
several schematic Penrose diagrams that summarize the 
kinds of behavior that have been observed in spherical
gravitational collapse of several Horndeski gravity theories.
All curvature singularities and sonic lines that have
been found in Horndeski gravitational collapse are spacelike, 
although given the difficulty of numerically evolving near those
points, new simulations may reveal that the boundary between
regular and irregular evolution may be null. 
A timelike boundary at a curvature singularity or sonic line
would lead to ill-posed evolution, unless suitable boundary
conditions could be found for that boundary region, or
if that region was excised along a null ray.

Some caveats about Fig.~\ref{fig:penrose_diagram_bh}:
during a numerical evolution, quasi-local definitions
of horizons (such as the apparent horizon) are typically
used to locate the surface of black holes\cite{Thornburg:2006zb}.
Black hole apparent horizons are located inside event
horizons so long as the Null Convergence Condition
($R_{\mu\nu}k^{\mu}k^{\nu}\geq0$ for all null vectors $k^{\mu}$)
holds\cite{hawking_ellis_1973}, but this condition
does not hold for many Horndeski gravity theories, even
within the weakly-coupled 
regime. For example the black hole
horizon can shrink with the growth of scalar
hair in $4\partial ST$ gravity\cite{Ripley:2019irj,Ripley:2019aqj,
Ripley:2020vpk,East:2020hgw,East:2021bqk}.
Because of this, there is generally no guarantee within Horndeski gravity that
an elliptic region contained within an apparent horizon will
remain within one for all times, although empirically
this does appear to be the case for solutions that 
are sufficiently weakly coupled exterior to all black 
hole horizons\cite{Ripley:2019irj,Ripley:2019aqj,
Ripley:2020vpk,East:2020hgw,East:2021bqk,Figueras:2020dzx,Figueras:2021abd}
(for a recent discussion about different notions of horizons in the
Horndeski gravity theories, see
\refcite{Reall:2021voz}).
Another subtlety is that the characteristic speeds of some degrees of freedom
in Horndeski gravity can travel faster than than the speed of light
as determined by the spacetime lightcones\cite{Akhoury:2011hr},
although in principle this can be accounted for by suitably
generalizing the notion of a horizon\cite{Reall:2021voz}.

Finally, we mention that order-reduced solutions to the Horndeski theories
only suffer from one potential problem not present in solutions
to GR: errors can grow secularly over time.
This can be a problem for accurately evolving binary black hole/neutron star
spacetimes over many orbits.
The issue of the \emph{secular growth} of errors is generic
to ordinary perturbation theory, and can be addressed using
\emph{renormalization group} methods, which have yet though
to be implemented in numerical relativity codes.

\subsection{Exact equations:
failure of hyperbolicity in the strongly-coupled regime
\label{sec:failure_strong_hyperbolicity}
}
Strong/weak hyperbolicity in the weakly-coupled regime is 
formulation-dependent and gauge-dependent.
In the strongly coupled regime, 
the hyperbolicity of some Horndeski theories can fail in a more
dramatic way: the equations of motion can develop fundamentally
elliptic degrees of freedom.
More formally, the equations of motion for many Horndeski gravity theories
form a system of \emph{mixed-type} 
partial differential equations.
The mixed-type property of the Horndeski equations of motion sets
them apart from the other PDE of classical physics: the
Einstein equations of GR, the Maxwell equations of
Electrodynamics, the Euler equations of fluid dynamics,
or the classical equations of motion for
the standard model\footnote{We note though in rotating frames of reference
that the classical wave equation becomes a mixed-type equation.
This behavior is not an \emph{inherent} property of the classical wave
equation though: it simply reflects a particular choice of coordinates.
For more discussion see Ref.~\refcite{Stewart_2001}}.
Those equations of motion for those
theories are always hyperbolic in a Lorentzian spacetime.

The model equations for mixed-type behavior are the 
Tricomi and Keldysh equations\cite{https://doi.org/10.1002/cpa.3160230404,chen2015tricomi}, 
which respectively are
\begin{align}
   \partial_x^2u
   +
   x
   \partial_y^2u
   =
   0
   ,\qquad
   \partial_x^2u
   +
   \frac{1}{x}
   \partial_y^2u
   =
   0
   .
\end{align}
These equations are hyperbolic when $x<0$, 
parabolic when $x=0$, and elliptic when $x>0$.
The characteristic speeds in the Keldysh equation
blow up before reaching the elliptic region; thus instabilities may appear 
in a code before reaching the elliptic region because of the 
Courant-Friedrich-Lewy condition being violated in explicit time-stepping codes.
By contrast, the characteristic speeds remain bounded
up until the formation of the elliptic region for the Tricomi equation.
The boundary between the elliptic and hyperbolic region of a mixed-type
equation is called the \emph{sonic line}\cite{rassias1990lecture,otway2015elliptic}.
We provide a brief, more general review of mixed-type equations in 
\ref{sec:mixed_type}.

Given that black holes in GR are expected to generically
have curvature singularities in their interior\cite{wald2010general},
it is generally expected that black hole solutions to the Horndeski gravity
theories will become strongly coupled deep enough inside the hole. 
In general agreement with this,
sonic lines have been found to occur inside and outside
of black hole horizons for several different
Horndeski theories;
whether the region appears inside or outside of a horizon depends
on the relative size of the Horndeski terms in the action,
and the strength of the spacetime curvature outside the black hole
horizon\cite{Ripley:2019hxt,Bernard:2019fjb,Ripley:2019irj,
Ripley:2019aqj,Ripley:2020vpk,Figueras:2020dzx,East:2021bqk}.
Black hole excision has been applied to evolve black holes in
4$\partial$ST gravity\cite{East:2020hgw,East:2021bqk} to
excise the elliptic region.
Black holes in Cubic Horndeski gravity theories have also been
evolved using puncture-like coordinates \emph{without} using
black hole excision, 
by adiabatically setting to zero the Horndeski coefficients
inside the black hole horizon\cite{Figueras:2020dzx,Figueras:2021abd}
(thus the theory in some sense adiabatically reduces to GR
inside the black hole)--we note that the authors refer to this
scheme also as a form of ``excision''.
Sonic lines have been found in solutions to quadratic Horndeski
gravity theories as 
well\cite{Bernard:2019fjb,Lara:2021piy,Bezares:2020wkn}.

Recent work by Barausse et. al. have suggests that the elliptic
regions found in the solution to some quadratic Horndeski
gravity theories could be removed with a suitable choice
of gauge\cite{Bezares:2021dma}, at least for spherically
symmetric evolution of the theory.
It seems unlikely  
that elliptic regions could be removed for solutions
to all Horndeski theories,
although there is no rigorous proof that is the case.

Mixed-type problems appear in, for example fluid 
mechanics\cite{doi:10.1142/S0219891604000081,ferrari1968transonic}, 
and even in some numerical relativity
applications (namely: when studying hyperbolic
equations in a rotating frame of reference)\cite{Stewart_2001}.
Given these previous applications, some numerical methods have
been developed and successfully employed to solve mixed-type PDE
across a sonic line\cite{AZIZ197655,doi:10.1137/0718047,Stewart_2001}.
In principle then, it may be possible to solve the Horndeski
equations of motion beyond the sonic line into the elliptic region.
These solutions are likely to be of little physical interest though,
as the formation of the sonic line essentially signals the breakdown
of the weak-coupling regime of the solution at hand.

\subsection{Exact equations: shock/caustic formation
\label{sec:shock_formation}
}
The model equation for shock formation is Burgers' 
equation\cite{bateman_burgers,BURGERS1948171}
\begin{align}
   \label{eq:burgers_eqn}
   \partial_tu
   +
   \frac{1}{2}\partial_x\left(u^2\right)
   =
   0
   .
\end{align}
This equation is \emph{genuinely nonlinear}, that is the
characteristic speed is a non-constant function of
the field $u$ itself\cite{whitham2011linear,
alcubierre2008introduction,evans2010partial,rezzolla2013relativistic}
(in this case, the characteristic speed $c=u$).
Similarly to Burgers' equation, the equations of motion for
some Horndeski theories 
can also be genuinely nonlinear\cite{Tanahashi:2017kgn}. 
Shock formation for smooth initial data has been studied in particular detail for
quadratic-Horndeski gravity 
theories (k-essence)\cite{Felder:2002sv,Babichev:2016hys,
Pasmatsiou:2017vcw,Bezares:2020wkn}.

We note that shock formation can occur even
in vacuum GR, namely ``gauge-shocks'' can form from smooth initial
data\cite{Alcubierre:1996su,Alcubierre:1997ee,Alcubierre:2002iq,Reimann:2004wp}.
Unlike in GR, shocks can form from smooth initial data for k-essence regardless
of the gauge/formulation used, at least
in $1+1$ dimensional setups\cite{Babichev:2016hys,Tanahashi:2017kgn}. 
It is likely that it will be more difficult 
to form shocks in $3+1$ asymptotically
flat evolution as compared to $1+1$ evolution 
due to the effect of dispersion,
although further work will be needed in this direction to determine
how prevalent shocks are in different subsets of Horndeski gravity.

Even when shocks do form,
in principle it is possible to (numerically) 
solve the Horndeski theories by formulating the shock front as a 
\emph{Riemann problem}\cite{2013numerical,rezzolla2013relativistic}, 
provided one writes the Horndeski equations of motion 
as a system of conservation laws, that is into a system 
of the form\cite{rezzolla2013relativistic}
\begin{align}
   \partial_0\left(\sqrt{h}\textbf{U}\right)
   +
   \partial_i\left(\sqrt{h}\textbf{F}^i\right)
   =
   \textbf{S}
   ,
\end{align}
where $\textbf{U}$ is the vector of conserved variables, $\textbf{F}^i$
are the vector of flux functions, $\textbf{S}$ is the vector of
source terms, and
we have used the $3+1$ decomposition \eqref{eq:3p1_decomposition}.
Once a set of equations has been written in conservative form,
a numerical method typically finds a \emph{weak solution} to solve
the equations through shocks.
Conservation laws can have multiple weak solutions; picking the ``physical''
one can require a detailed knowledge of the PDE properties
of the equations of motion. As one example, for the Burgers
equation (\eqref{eq:burgers_eqn}; or more generally the Euler
equations of fluid dynamics) one can introduce an
\emph{entropy condition}, and the physical weak solution is then
the entropy increasing solution\cite{2013numerical}.

Some work has been done to successfully numerically evolve the 
equations of motion for some quadratic Horndeski (K-essence) theories 
using shock-capturing methods, including in the inspiral
and collision of binary neutron stars with k-essence with the
equations of motion written in a conservative form\cite{Bezares:2020wkn,
terHaar:2020xxb,Bezares:2021yek,Bezares:2021dma}.
This being said, other Horndeski theories may suffer a breakdown
in hyperbolicity 
(see Sec.~\ref{sec:failure_strong_hyperbolicity}),
before a shock can form.
The study of weak solutions to the Horndeski equations has not been
thoroughly studied.

\subsection{Perturbative solutions: secular instabilities, 
and the numerical dynamical renormalization group
\label{sec:perturbative_secular_instabilities}
}
The main technical challenge to using naive perturbation theory
(Sec.~\ref{sec:perturbative_well_posedness})
for long-time evolution
is that in this approach, 
errors can grow secularly in time\cite{bender2013advanced}.
The problem can be illustrated with the following perturbed
simple harmonic oscillator:
\begin{align}
   \frac{d^2 u}{dt^2} + \omega^2 u + \epsilon \frac{d^2u}{dt^2}
   =
   0
   ,\qquad
   u(0) = 0
   ,\;\;
   u'(0) = \omega
   .
\end{align}
Here $\omega$ is a constant and $\epsilon\ll1$.
The solutions order-by-order in $\epsilon$ are
\begin{subequations}
\begin{align}
   u^{(0)}(t)
   &=
   \sin\left(\omega t\right)
   ,\\
   u^{(1)}(t)
   &=
   \frac{1}{2}\epsilon\left(
      \sin\left(\omega t\right)
      -
      t\omega\cos\left(\omega t\right)
   \right)
   ,\\
   \vdots\nonumber
\end{align}
\end{subequations}
The presence of $t\omega \cos\left(\omega t\right)$ signals
the existence of a secular instability (the full
solution is $u(t) = \left(1+\epsilon\right)
\sin\left(\omega t/\left(1+\epsilon\right)\right)$).
Solving to higher orders in $\epsilon$ perturbatively will only
introduce higher powers of $t$ in the solution.

Several methods have been developed to address this secular growth of
error (such as the method of multiple 
scales\cite{bender2013advanced,kevorkian2012multiple}); 
most of these method have been shown to be special cases
of a more general method call
the \emph{dynamical renormalization group} (DRG)\cite{Chen:1995ena,
Kunihiro:1995zt,10.1143/PTPS.99.244,Ei:1999pk}.
Operationally speaking, the essence of the DRG is promote
the constants of the background solution to dynamical functions,
which are chosen to cancel out the secularly growing terms
in the solution to a given perturbative order.
In the case of the perturbed simple harmonic oscillator,
the amplitude $A$ and frequency $\omega$ of the zeroeth order solution 
would be promoted to a function of $t$, and would be corrected. 
In this simple case, the constant shifts 
$A\to1 + \epsilon, \omega \to \omega - \epsilon\omega $ would suffice
to cancel out the secular growth to linear order in $\epsilon$.

The DRG has recently been reformulated 
by Galvez-Ghersi and Stein\cite{GalvezGhersi:2021sxs},
which they call the \emph{numerical dynamical
renormalization group}.
This is because their approach allows for the renormalization
of parameters for theories whose (background) 
solution may only be found numerically.
We note that the application of the DRG formally requires the full
solution to be described by 
a finite dimensional attractor manifold in parameter space.
The Einstein (and Horndeski) equations of motion are partial differential
equations, which formally have an infinite number of control parameters
(for example, one needs to set $\phi,\partial_t\phi$ at each spatial point
on the initial data slice).
Because of this, it is not
immediately obvious how the numerical DRG can be applied to generic
solutions to these theories.

At least in the context of binary black hole mergers though, 
there is reason to believe the DRG could prove to be useful.
From Post-Newtonian theory, it is known
that the problem of binary black hole/neutron star inspiral
can be very accurately described using a finite number of 
parameters; for a review see 
Ref.~\refcite{Blanchet:2013haa}\footnote{\emph{Surrogate modeling} 
of gravitational waveforms 
also shows that gravitational waveforms for a variety of gravitational wave
sources can be accurately described using a relatively small 
number of parameters\cite{Tiglio:2021ysj}.}.
Thus one expects there is a finite-dimensional attractor manifold
for the two-body problem in GR (and Horndeski gravity, and other
modified theories of gravity).
In principle then the numerical DRG should 
be able to control the secular divergences
encountered in the modeling of gravitational waves from black holes/neutron
inspiral for a variety of modified gravity theories, including potentially
perturbative solutions to the Horndeski theories of gravity.
In particular,
the numerical DRG has been proposed to be applied directly to the 
gravitational waveforms
computed from perturbative solutions to 
modified gravity theories\cite{GalvezGhersi:2021sxs}.

\section{Constructing exact initial data
\label{sec:constructing_initial_data}
}

As in GR, the equations of motion for the
Horndeski theories form an overdetermined
system of partial differential equations,
and there are a set of constraint equations that must be satisfied
on the initial data surface.
In this section we will focus only on the exact constraint equations;
the perturbative equations of motion 
(Sec.~\ref{sec:perturbative_well_posedness}) reduce to those
of the Einstein equations plus lower order terms, and thus can
be solved using the same techniques already used in numerical
relativity\cite{Cook:2000vr}.

Let $\left(M,g\right)$ be a smooth, 4-dimensional, globally hyperbolic
spacetime. An initial data set is the set 
$\left(\Sigma,h_{ij},K_{ij},\phi,\partial_0\phi\right)$, 
where $\Sigma$ is a smooth 3-dimensional spacelike submanifold
of $M$, $h_{ij}$ and $K_{ij}$ are
the Riemannian metric and extrinsic curvature induced on $\Sigma$,
and $\phi$ and $\partial_0\phi$ are the scalar field and its first time
derivative on $\Sigma$.
The indices $i,j$ run over the spatial tensor components induced
on $\Sigma$.
We denote the induced metric-compatible covariant derivative
operator on the spatial hypersurface with $D_i$, and the
induced Christoffel symbols and tensors with a ${}^{(3)}$ superscript.

The constraint equations can be cast into a generalization
of the \emph{Hamiltonian} and \emph{momentum} constraints of GR.
For a given $\Sigma$, we let $n^{\mu}$ be the unit timelike future-directed
null vector orthogonal to $\Sigma$.
We can then write the spacetime metric as 
\begin{align}
   \label{eq:covariant_3p1_decomp}
   g_{\mu\nu}
   =
   -
   n_{\mu}n_{\nu}
   +
   h_{\mu\nu}
   .
\end{align}
The induced form of the metric $h_{\mu\nu}$ on $\Sigma$ is $h_{ij}$.
The covariant expression for the extrinsic curvature is
\begin{align}
   K_{\mu\nu}
   \equiv
   -
   h_{\mu}{}^{\alpha}h_{\nu}{}^{\beta}\nabla_{\alpha}n_{\beta}
   .
\end{align}
The generalized Hamiltonian and momentum constraints are defined to be 
\begin{align}
   \label{eq:general_hamiltonian_constraint}
   \mathcal{H}
   \equiv
   E^{(g)}_{\gamma\delta}n^{\gamma}n^{\delta}
   =
   0
   ,\\
   \label{eq:general_momentum_constraint}
   M_{\mu}
   \equiv
   E^{(g)}_{\gamma\delta}n^{\gamma}h^{\delta}_{\mu}
   =
   0
   .
\end{align}

There are two main challenges to constructing initial data:
\begin{enumerate}
   \item Casting the constraint equations into a form that
      allows for a well-posed boundary value problem.
   
   \item Relating the evolution variables ($g_{\mu\nu},\phi$)
      to physically meaningful quantities, to construct
      astrophysically realistic initial data.
\end{enumerate}
A large number of techniques have been developed over the years to address
these two problems in GR (for reviews, see
Refs.~\refcite{alcubierre2008introduction,baumgarte2010numerical,Cook:2000vr}).
We review how some of these techniques can be extended to
the problem of constructing
constraint satisfying initial data for Horndeski gravity theories.
In particular,
in Sec.~\ref{eq:elliptic_character_constraints} we show that
provided the Horndeski corrections are weakly coupled 
(for a review of this terminology see Sec.~\ref{sec:weak_coupling}),
the constraint equations can be shown to be elliptic
under the conformal transverse-traceless decomposition, and
in Sec.~\ref{sec:puncture_initial_data} 
we show how black hole ``puncture'' initial data has been 
extended to some Horndeski gravity theories\cite{Kovacs:2021lgk}.
There has been essentially no numerical work on constructing numerical
solvers to the general constraint equations in Horndeski gravity.

\subsection{The constraint equations: general properties
\label{eq:constraint_equations_general_properties}
}
We pick ADM-like coordinates adapted to the spacelike hypersurface
$\Sigma$, so that $x^0$ is the timelike coordinate:
\begin{align}
   \label{eq:3p1_decomposition}
   g_{\mu\nu}dx^{\mu}dx^{\nu}
   =
   -
   N^2\left(dx^0\right)^2
   +
   h_{ij}\left(N^idx^0 + dx^i\right)\left(N^jdx^0 + dx^j\right)
   .
\end{align}
The unit orthogonal timelike vector is then $n_{\alpha}=\left(-N,0,0,0\right)$.
The lapse is $N$ and the shift vector is $N^i$.

We first show that 
$\partial_0^2\phi$ and $\partial_0^2g_{\mu\nu}$
do not appear in the Horndeski constraint equations.
From the constraint equations 
\eqref{eq:general_hamiltonian_constraint} 
\eqref{eq:general_momentum_constraint}
and tensor equations of motion \eqref{eq:tensor_eom_horndeski},
we see that the presence of second
time derivatives comes from there being repeated $0$ components
in contractions with the generalized Kronecker delta tensor:
\begin{align}
   &
   n_{\alpha}n^{\beta}
   \delta^{\alpha\gamma_1\gamma_2}_{\beta\delta_1\delta_2}
   &
   n_{\alpha}h_{\mu}^{\beta}
   \delta^{\alpha\gamma_1\gamma_2}_{\beta\delta_1\delta_2}
   \nonumber\\
   &
   n_{\alpha}n^{\beta}
   \delta^{\alpha\gamma_1\gamma_2\gamma_3}_{\beta\delta_1\delta_2\delta_3}
   &
   n_{\alpha}h_{\mu}^{\beta}
   \delta^{\alpha\gamma_1\gamma_2\gamma_3}_{\beta\delta_1\delta_2\delta_3}
   \nonumber\\
   &
   n_{\alpha}n^{\beta}
   \delta^{\alpha\gamma_1\gamma_2\gamma_3}_{\beta\delta_1\delta_2\delta_3}
   &
   n_{\alpha}h_{\mu}^{\beta}
   \delta^{\alpha\gamma_1\gamma_2\gamma_3}_{\beta\delta_1\delta_2\delta_3}
   \nonumber
\end{align}
From the contractions of the Riemann tensor 
(see \ref{sec:spatial_conformal_decomposition}),
in adapted coordinates
we see that the only two time derivative terms that could act on the
metric would come from the contraction 
$
h_{\mu_1}{}^{\gamma_1}
n^{\gamma_2}
h^{\nu_1}{}_{\delta_1}
n_{\delta_2}
R_{\gamma_1\gamma_2}{}^{\delta_1\delta_2}
$.
From the contractions on $\nabla_{\gamma}\nabla^{\delta}\phi$
(see \ref{sec:spatial_conformal_decomposition}),
in adapted coordinates we see that the only two
time derivative terms could come from contractions on two $n^{\mu}$ vectors
(the derivatives $D$ are purely spatial in the induced coordinates):
$n^{\mu}n_{\nu}\nabla_{\mu}\nabla^{\nu}\phi$.
Our task then is to show that there are no repeated $n^{\mu}$ vectors 
in the constraints with a raised/lowered index (for example there is no
contraction like $n_{\alpha}n_{\gamma_1}$ in the constraints).
By the antisymmetry of the Kronecker delta tensors we easily see this is
the case. For example we have 
\begin{align}
   n_{\alpha}h_{\mu}^{\beta}
   \delta^{\alpha\gamma_1\gamma_2}_{\beta\delta_1\delta_2}
   \times
   n_{\gamma_1}n^{\delta_1}
   h_{\gamma_2}^{\kappa_2}
   h_{\rho_2}^{\delta_2}
   \times
   \cdots
   =
   0
   .
\end{align}
Using a similar argument it is easy to show that
there are no terms multiplying $\partial_0^2\phi$ or 
$\partial_0^2g_{\mu\nu}$ in the constraint equations in
$4\partial ST$ (which we know must also
be true as it is a special case of a Horndeski gravity 
theory)\cite{East:2020hgw}.

We provide a more general argument that shows that
there are no terms with second order time derivatives
in the constraint equations for Horndeski gravity--which can be be 
straightforwardly generalized to scalar-tensor
theories with multiple scalar fields\cite{Tanahashi:2017kgn,Kobayashi:2019hrl},
vector tensor theories\cite{doi:10.1063/1.522837,Davies:2021frz},
or more generally any tensor theory that has second order equations of motion
that can be derived from an action--in 
\ref{sec:general_properties_principal_symbol}.

As a corollary of there being no second time derivatives in the constraint
equations, we see that if we set
\begin{align}
\label{eq:general_vacuum_id_horndeski}
   \phi\big|_{\Sigma} 
   = 
   \partial_0\phi\big|_{\Sigma} 
   = 
   0
   ,
\end{align}
then the constraint equations for Horndeski gravity reduce to those of 
vacuum GR. This follows because from 
Eq.~\eqref{eq:general_vacuum_id_horndeski} as it implies 
$\partial_{\mu}\phi=0$
and $\partial_{\mu}\partial_i\phi=0$, and we have already
established there are no terms like $\partial_0^2\phi$ in the constraint
equations.
This result has also been shown to be true for 
$4\partial ST$ theory\cite{East:2020hgw} 
(as before, we can alternatively note that these
results hold for $4\partial ST$ theory as it is a Horndeski theory).

An additional corollary of the fact that there are no repeated
time derivatives acting on the metric or scalar degrees
of freedom is that the constraint equations can be formulated as 
a system of elliptic PDE when the Horndeski terms are weakly coupled.
We sketch an argument that explicitly shows this to be true
in a conformal transverse-traceless decomposition
in Sec.~\ref{eq:elliptic_character_constraints}.
\subsection{Example: the elliptic character of the exact constraint equations
   under the conformal transverse-traceless (CTT) decomposition 
\label{eq:elliptic_character_constraints}
}
Here we show in more detail how the constraints can be written as
a set of elliptic equations for weakly coupled solutions
by working with the conformal
transverse-traceless (CTT) decomposition\cite{1979sgrr.work...83Y}.
We will only consider the principal part of the
constraint equations, as it is the part that determines the
character of the constraints as a set of PDE.
Kovacs\cite{Kovacs:2021lgk} has shown that for $4\partial ST$ gravity, 
for a weakly coupled solution under a 
CTT decomposition (and for a conformal thin sandwich decomposition)
the constraints formed an elliptic set of PDE. 
Moreover, he showed that a unique solution exists to the elliptic boundary
value problems on asymptotically Euclidean initial slices 
under similar conditions as in the case of General Relativity.
In demonstrating the fundamental ellipticity of the constraint equations in
the CTT formalism for the general constraint equations, this section 
demonstrates the plausibility of 
extending Kovacs proof to initial data for general Horndeski theories.
We emphasize that other decompositions commonly used in numerical relativity
could also likely lead to an elliptic set of constraint equations
for the Horndeski gravity theories for weakly-coupled solutions.

The conformal (Lichnerowicz-York) decomposition is:
\begin{align}
\label{eq:conformal_decomp_1}
   h_{ij}
   &=
   \psi^4\tilde{h}_{ij}
   ,\\
   K_{ij}
   &=
   \frac{1}{\psi^2}\tilde{A}_{ij}
   +
   \frac{1}{3}\gamma_{ij}K
   ,
\end{align}
where $\tilde{A}_{ij}$ is traceless.
Finally we decompose $\tilde{A}_{ij}$ 
with a transverse decomposition
\begin{align}
\label{eq:conformal_decomp_2}
   \tilde{A}_{ij}
   =&
   \left(\tilde{L}W\right)_{ij}
   +
   \tilde{A}_{(TT)ij}
   ,\\
   \left(\tilde{L}W\right)_{ij}
   \equiv&
   \tilde{D}_iW_j
   +
   \tilde{D}_jW_i
   -
   \frac{2}{3}\tilde{\gamma}_{ij}\tilde{D}_kW^k
   ,
\end{align}
where $\tilde{L}$ is the \emph{longitudinal operator}
(also called the \emph{vector gradient}) and 
$\tilde{A}_{(TT)ij}$ is a transverse-traceless tensor:
$\tilde{\gamma}^{ij}\tilde{A}_{(TT)ij}=\tilde{D}^i\tilde{A}_{(TT)ij}=0$.
As is the case in GR with a minimally coupled scalar field, 
the free degrees of freedom are
\begin{align}
   \tilde{\gamma}_{ij}
   ,\qquad
   K
   ,\qquad
   \tilde{A}_{(TT)ij}
   ,\qquad
   \phi
   ,\qquad
   \partial_0\phi
   ,
\end{align}
and the constraints will determine the degrees of freedom 
\begin{align}
   \psi
   ,\qquad
   W_i
   .
\end{align}

We first study the Hamiltonian constraint:
\begin{align}
\label{eq:intermediate_principal_part_ham_ctt}
   \mathcal{H}
   =&
   -
   \frac{1}{4}\left(
      1 
      + 
      \mathcal{G}_4 
      - 
      2X\partial_X\mathcal{G}_4 
      + 
      X \partial_{\phi}\mathcal{G}_5
   \right)
   n_{\alpha}n^{\beta}
   \delta^{\alpha\gamma_1\gamma_2}_{\beta\delta_1\delta_2}
   R_{\gamma_1\gamma_2}{}^{\delta_1\delta_2}
   \nonumber\\
   &+
   \frac{1}{4}\left(
      \partial_X\mathcal{G}_4
      -
      \partial_{\phi}\mathcal{G}_5
   \right)
   n_{\alpha}n^{\beta}
   \delta^{\alpha\gamma_1\gamma_2\gamma_3}_{\beta\delta_1\delta_2\delta_3}
   \nabla_{\gamma_1}\phi\nabla^{\delta_1}\phi
   R_{\gamma_2\gamma_3}{}^{\delta_2\delta_3}
   \nonumber\\
   &-
   \frac{1}{4}
   \left(X\partial_X\mathcal{G}_5\right)
   n_{\alpha}n^{\beta}
   \delta^{\alpha\gamma_1\gamma_2\gamma_3}_{\beta\delta_1\delta_2\delta_3}
   \nabla_{\gamma_1}\nabla^{\delta_1}\phi
   R_{\gamma_2\gamma_3}{}^{\delta_2\delta_3}
   \nonumber\\
   &+
   l.o.t.
\end{align}
Given the antisymmetry of the Kronecker delta tensor, we see
the only nonzero components of the Riemann tensor will come
from contractions on $h_{\mu\nu}$.
As $K_{\mu\nu}$ contains only first derivatives of the metric,
we only focus on ${}^{(3)}R_{\mu_1\mu_2}{}^{\nu_1\nu_2}$:
$\mathtt{P}\left[R_{\gamma_1\gamma_2}{}^{\delta_1\delta_2}\right] 
\to
\mathtt{P}\left[{}^{(3)}R_{i_1i_2}{}^{j_1j_2}\right]
$,
and $\delta^{i\cdots}_{j\cdots}$ are the Kronecker delta tensors
induced on the spatial surface $\Sigma$.
Continuing to work in adapted coordinates, 
under the conformal decomposition we see that
(see \ref{sec:spatial_conformal_decomposition})
\begin{align}
   {}^{(3)}R_{i_1i_2}{}^{j_1j_2}
   =
   -
   8
   \frac{1}{\psi^5}
   \delta^{[j_1}_{[i_1}\tilde{D}_{i_2]}\tilde{D}^{j_2]}\psi
   +
   l.o.t.
\end{align}
Plugging this in to 
Eq.~\eqref{eq:intermediate_principal_part_ham_ctt},
we see that the principal part of Hamiltonian constraint
for the conformal factor is
\begin{align}
   \mathcal{H}
   =&
   \tilde{\Delta}\psi
   +
   F\left(\partial_i\partial_j\psi\right)
   +
   l.o.t.
\end{align}
where the $F$ contains terms that have second derivatives acting
on $\psi$, but which are suppressed in the weakly-coupled limit.
Provided then that the Horndeski terms are sufficiently weakly coupled 
(Eq.~\eqref{eq:weak_coupling_conditions}), 
this is an elliptic equation for $\psi$ as the
Laplacian $\tilde{\Delta}\equiv\tilde{D}_i\tilde{D}^i$ is 
elliptic. This follows as the principal symbol for the Laplacian is symmetric
positive definite matrix, and for a sufficiently 
small symmetric perturbation,
a symmetric positive definite matrix remains positive definite.

We next turn to the momentum constraint:
\begin{align}
\label{eq:intermediate_principal_part_mom_ctt}
   \mathcal{M}_{\mu}
   =&
   -
   \frac{1}{4}\left(
      1 
      + 
      \mathcal{G}_4 
      - 
      2X\partial_X\mathcal{G}_4 
      + 
      X \partial_{\phi}\mathcal{G}_5
   \right)
   n_{\alpha}h_{\mu}^{\beta}
   \delta^{\alpha\gamma_1\gamma_2}_{\beta\delta_1\delta_2}
   R_{\gamma_1\gamma_2}{}^{\delta_1\delta_2}
   \nonumber\\
   &+
   \frac{1}{4}\left(
      \partial_X\mathcal{G}_4
      -
      \partial_{\phi}\mathcal{G}_5
   \right)
   n_{\alpha}h_{\mu}^{\beta}
   \delta^{\alpha\gamma_1\gamma_2\gamma_3}_{\beta\delta_1\delta_2\delta_3}
   \nabla_{\gamma_1}\phi\nabla^{\delta_1}\phi
   R_{\gamma_2\gamma_3}{}^{\delta_2\delta_3}
   \nonumber\\
   &-
   \frac{1}{4}
   \left(X\partial_X\mathcal{G}_5\right)
   n_{\alpha}h_{\mu}^{\beta}
   \delta^{\alpha\gamma_1\gamma_2\gamma_3}_{\beta\delta_1\delta_2\delta_3}
   \nabla_{\gamma_1}\nabla^{\delta_1}\phi
   R_{\gamma_2\gamma_3}{}^{\delta_2\delta_3}
   \nonumber\\
   &+
   l.o.t.
   .
\end{align}
Given the antisymmetry of the Kronecker delta, the only nonzero contractions
against the Riemann tensor will come from
$   
h_{\mu_1}{}^{\gamma_1}
h_{\mu_2}{}^{\gamma_2}
h^{\nu_1}{}_{\delta_1}
h_{\nu_2}{}^{\delta_2}
R_{\gamma_1\gamma_2}{}^{\delta_1\delta_2}
,
$
and
$   
h_{\mu_1}{}^{\gamma_1}
h_{\mu_2}{}^{\gamma_2}
h^{\nu_1}{}_{\delta_1}
n_{\delta_1}
R_{\gamma_1\gamma_2}{}^{\delta_1\delta_2}
.
$
From the latter contraction we see the principal part will come
from $\nabla K$ terms; in adapted coordinates we have 
\begin{align}
   D^{[j_2}K^{j_1]}_{i_3}
   =
   \frac{1}{\psi^{10}}\tilde{D}^{[j_2}\left(\tilde{L}W\right)^{j_1]}_{i_3}
   +
   l.o.t.
\end{align}
From $h^{\mu}_{\nu}n^{\nu}=0$, we see that in adapted coordinates
the momentum constraint is
\begin{align}
\label{eq:principal_part_mom_ctt}
   \mathcal{M}^{i}
   =&
   \left(\tilde{\Delta}_LW\right)^i
   +
   F^i\left(\partial_j\partial_kW^l,\partial_k\partial_k\psi\right)
   +
   l.o.t.
\end{align}
where $F^i$ contains terms that have second derivatives acting on
$W^l$ and $\psi$, but which are suppressed in the weak-coupling limit.
While there are now second derivatives of $\psi$ in the momentum
constraint, provided we work in the weak-coupling limit, the
\emph{system} of PDE formed by the Hamiltonian and momentum constraints
remain fundamentally elliptic in character as the vector Laplacian
$\left(\tilde{\Delta}_LW\right)^i
\equiv 
\tilde{D}_j\left(\tilde{L}W\right)^{ij}$ 
is elliptic.

While we have shown that the constraint equations are fundamentally
elliptic in Horndeski gravity, we reiterate that the well-posedness
of asymptotically flat solutions to
the constraint equations as a set of elliptic PDE has \emph{not} yet been
proven for general weakly-coupled Horndeski gravity theories.
Such a proof has only recently been constructed  
by Kovacs\cite{Kovacs:2021lgk} for $4\partial ST$ gravity, 
which contains a particular subset of the Horndeski theory terms.
We expect though that a proof similar to his should follow for general Horndeski
gravity theories, as his argument essentially rested on the fundamentally
elliptic nature to the constraint PDE for weakly-coupled solutions,
and the fact that nonlinear lower order
terms generally do not control the solution properties of sufficiently regular
asymptotically flat solutions.

\subsection{Example:
Bowen-York ``puncture'' initial data in $4\partial ST$ gravity
\label{sec:puncture_initial_data}
}
Kovacs has proposed a prescription to construct Bowen-York 
``black hole puncture'' 
initial data\cite{Kovacs:2021lgk} in $4\partial ST$ gravity\footnote{We 
restrict ourselves to $4\partial ST$ gravity as--at least
as of the publication of this review--there is
no general proposal for black hole puncture initial data for
general Horndeski gravity theories.}
This class of initial data has been widely used in GR
as it transparently allows one to set the 
approximate mass, momentum, and spin
of multiple black hole solutions\cite{Ansorg:2004ds}.

Prescribing Bowen-York initial data in GR\cite{PhysRevD.21.2047,Brandt:1997tf} 
involves setting 
\begin{align}
   \tilde{h}_{ij}
   =
   \delta_{ij}
   ,\qquad
   K
   =
   0
   ,\qquad
   \tilde{A}_{(TT)}^{ij}
   =
   0
   .
\end{align}
With the above conditions,
in general relativity the momentum constraint equation can be solved with
\begin{align}
   W^i
   =
   -
   \frac{1}{4r}\left(7P^i + P^j\hat{x}_j \hat{x}^i\right)
   -
   \frac{1}{r^2}\epsilon^{ijk}S_j\hat{x}_k
   ,
\end{align}
where $x^i$ are the Euclidean coordinates in $\mathbb{R}^3$,
$r=\sqrt{\delta_{ij}x^ix^j}$ is the Euclidean distance from the puncture,
$\hat{x}^i/r$ is the unit Euclidean vector, and $\epsilon^{ijk}$
is the Levi-Cevita symbol.
The free initial data are the vectors $P^i,S^i$.
The vector $P^i$ corresponds to the total ADM linear momentum
of the spacetime, and $S^i$ is the total ADM angular momentum 
of the spacetime\cite{PhysRevD.21.2047}, that is with these choices we have 
\begin{subequations}
\label{eq:adm_momenta_puncture_gr}
   \begin{align}
   P_{ADM}^i
   &\equiv
   \frac{1}{8\pi}\lim_{r\to\infty}\int_{\mathbb{S}_2} dA\left(
      K^{ij}
      -
      K
      h^{ij}
   \right)
   \hat{x}_j
   =
   P^i
   ,\\
   S^i_{ADM}
   &\equiv
   \frac{1}{8\pi}\lim_{r\to\infty}\int_{\mathbb{S}_2} dA\left(
      K_{jk}
      -
      K
      h_{jk}
   \right)
   \epsilon^{ijl}x_l\hat{x}^k
   =
   S^i
   .
\end{align}
\end{subequations}
While these are global quantities, it turns out they are typically
close to the quasi-local definitions of the black hole linear
and angular momentum. The remainder of the momentum is carried
in ``junk radiation'', the amount of which is typically dependent on
the initial black hole spin.
As the momentum constraint is linear in $W^i$, this initial data can 
be superposed for a collections of black holes
\begin{align}
   W^i
   =
   -
   \sum_n
      \left(
         \frac{1}{4r_{(n)}}\left(
            7\left(P^{(n)}\right)^i 
            + 
            \left(P^{(n)}\right)^j\hat{x}_j \hat{x}^i
         \right)
         +
         \frac{1}{r_{(n)}^2}\epsilon^{ijk}\left(S^{(n)}\right)_j\hat{x}_k
      \right)
   ,
\end{align}
where $r_{(n)}\equiv\left|{\bf x} - {\bf c}_{(n)}\right|$, the
Euclidean distance from the $n^{th}$ puncture
(${\bf c}_{(n)}$ are at the black hole puncture locations).
With puncture initial data, we still need to solve the Hamiltonian
constraint numerically, which forms an elliptic equation for $\psi$.
This can be solved more easily by making the following field
redefinition\cite{Brandt:1997tf}
\begin{align}
   \psi
   =
   1
   +
   \frac{1}{\mu}
   +
   u
   ,\qquad
   \frac{1}{\mu}
   \equiv
   \sum_{n}\frac{m_{(n)}}{2r_{(n)}}
   ,
\end{align}
That is, we analytically remove the ``singular'' 
term from the conformal factor.
This renders the solution $\psi$ to be regular 
(more precisely, $C^2$; see \refcite{Brandt:1997tf})
throughout $\mathbb{R}^3$ for asymptotically flat boundary conditions.

Kovacs\cite{Kovacs:2021lgk} has generalized this
initial data for $4\partial ST$ gravity. While this only contains
a particular subset of Horndeski gravity, it does contain the term
that causes scalar ``hairy'' black holes (namely, the
Gauss-Bonnet coupling $\beta\left(\phi\right)\mathcal{G}$),
so it is perhaps the most interesting Horndeski theory to study
in the context of binary black hole systems.
His construction relies on the linearity of the momentum
constraint for the theory when it is written in terms of the
canonical conjugate momentum for the tensor and scalar
degree of freedom\footnote{We note
that while the relation between the canonical momenta
$\pi^{ij}$, $\pi_{\phi}$
and $K_{ij}$, $\partial_t\phi$ for $4\partial$ST gravity
is nonlinear\cite{Julie:2020vov}, for weakly-coupled solutions it is
possible to relate the two by iteratively solving the
relations 
Eq.~\eqref{eq:canonical_momenta_4dst}\cite{Kovacs:2021lgk}.}
\cite{Julie:2020vov,Kovacs:2021lgk}
\begin{align}
\label{eq:canonical_momenta_4dst}
   \pi^{ij}
   \equiv
   \frac{\delta\mathcal{L}}{\delta\left(\partial_0h^{ij}\right)}
   ,\qquad
   \pi_{\phi}
   \equiv
   \frac{\delta\mathcal{L}}{\delta\left(\partial_0\phi\right)}
   .
\end{align}
Kovacs' prescription for puncture initial data is to
then decompose $\pi^{ij}$ and $\pi_{\phi}$ in the following way: 
\begin{align}
   \tilde{\pi}_{ij}
   \equiv
   \psi^2\left(
      \frac{1}{\sqrt{h}}\pi_{ij}
      -
      \frac{1}{3}h^{kl}\pi_{kl}
      h_{ij}
   \right)
   ,\qquad
   \tilde{\pi}_{\phi}
   \equiv
   \psi^6\frac{\pi_{\phi}}{\sqrt{h}}
   .
\end{align}
Then the trace is set to zero, and the tracefree part is
set to the same functional form as the tracefree part of the extrinsic
curvature is in GR:
\begin{subequations}
\begin{align}
   \tilde{h}_{ij}
   &=
   \delta_{ij}
   ,\\
   \tilde{\pi}_{\phi}
   &=
   0
   ,\\
   \tilde{\pi}_{ij}
   &=
   \partial_iW_j
   +
   \partial_jW_i
   -
   \frac{2}{3}\delta_{ij}\partial_kW^k
   ,\\
   W^i
   &=
   -
   \sum_n
      \left(
         \frac{1}{4r_{(n)}}\left(
            7\left(P^{(n)}\right)^i 
            + 
            \left(P^{(n)}\right)^j\hat{x}_j \hat{x}^i
         \right)
         +
         \frac{1}{r_{(n)}^2}\epsilon^{ijk}\left(S^{(n)}\right)_j\hat{x}_k
      \right)
   .
\end{align}
\end{subequations}
As the canonical momentum reduces to the extrinsic curvature in the
asymptotically flat limit ($r\to\infty$), it is straightforward to
show that the spacetime linear and angular momentum are set by
the total $P^i$ and $S^i$ 
(see Eq.~\eqref{eq:adm_momenta_puncture_gr})\cite{Kovacs:2021lgk}.
We have also defined a traceless and conformally rescaled
conjugate momentum.
In this setup, 
the fields $\phi,\partial_0\phi$ can be freely specified so long
as they satisfy $\tilde{\pi}_{\phi}=0$.
This initial data satisfies the momentum constraint in 
$4\partial ST$ gravity, and in the weak-coupling limit
the conformal factor $\psi$ obeys an elliptic equation of motion
(the Hamiltonian constraint).

\section{Evolution of compact objects in 
Horndeski gravity theories: a brief survey of numerical work
\label{sec:general_survey_numerical_work}}

\subsection{Bergmann--Wagoner scalar-tensor theories
\label{sec:scalar_tensor_special_case_numerical}}
   First we survey the numerical evolution of ``scalar-tensor'' gravity
theories (see also Sec.~\ref{sec:scalar_tensor_special_case}).
Most numerical work for these theories
has been in the Einstein frame,
where there is no scalar-tensor coupling in the principal part
of the equations of motion.
In that frame, from a numerical standpoint evolving the equations
of motion for scalar-tensor theories is no more difficult than
evolving the equations of motion for minimally a coupled scalar field.
Due to the no-hair theorems for black holes in these 
theories\cite{Hawking:1972qk,Herdeiro:2015waa},
numerical work on this theory has focused on the evolution
of (binary) neutron stars, which can exhibit interesting solutions such as 
\emph{spontaneous scalarization}\cite{Damour:1996ke}.
Some work on black holes in these theories
has been performed in spherical symmetry as 
well\cite{Scheel:1994yr,Scheel:1994yn}.
In particular, much of the numerical work on scalar-tensor
theories has focused on stellar collapse of
a spherically symmetric scalarized star to a black hole, as in the process
of forming a black hole the star must shed a large amount
of scalar hair, which could be potentially observable
through measurement of a scalar polarization mode in gravitational 
waves\cite{Novak:1999jg,
Gerosa:2016fri,
Sperhake:2017itk,
Cheong:2018gzn,
Rosca-Mead:2019seq,Rosca-Mead:2020ehn,Geng:2020slq,
Huang:2021tpu,Kuan:2021yih,
Kuan:2022oxs}.
Other work in spherical symmetry has been done on scalarized
stars\cite{Mendes:2021fon}, including studies
of potential \emph{screening} 
effects\cite{Khoury:2003aq,Khoury:2003rn,Joyce:2014kja,Quiros:2019ktw} 
of these theories\cite{Dima:2021pwx}. 
Relatively little work on full $3+1$ collisions 
of scalarized stars has been done for scalar-tensor gravity theories.
\subsection{Quadratic Horndeski gravity
\label{sec:quadratic_horndeski_numerical}}
The quadratic Horndeski theories have been used to construct models of 
dark energy/inflation\cite{Armendariz-Picon:2000nqq,
Armendariz-Picon:2000ulo} 
that have not been observationally 
ruled out\cite{Baker:2017hug,Creminelli:2018xsv,Creminelli:2019kjy}.
Moreover solutions to 
variants of the theory can 
exhibit a screening mechanism\cite{Joyce:2014kja} called
\emph{k-mouflage}\cite{Babichev:2009ee}, whereby
the spacetime geometry around a scalarized star 
can closely resemble the vacuum exterior of an unscalarized star.
Numerical work on the quadratic gravity theories has focused
mostly on investigating the well-posedness of these theories
in the strong field, dynamical 
regime\cite{Bernard:2019fjb,Bezares:2020wkn,Lara:2021piy}.
In particular, there has been work on investigating the
formation of shocks and sonic lines in these theories,
and the dynamics of the theory during the onset of
k-mouflage\cite{terHaar:2020xxb,
Bezares:2021yek,Dima:2021pwx}, 
which can move the solution to the strongly-coupled regime. 
Asymptotically flat black holes cannot support stable scalar hair 
(outside of supper-radiance)\cite{Graham:2014mda},
so much of the attention has been focused on (neutron) star solutions.

More recently, Bezares et. al.\cite{Bezares:2021dma} have performed
full $3+1$ evolution of binary neutron star mergers in a variant
of quadratic gravity that was predicted to lead to screening.
To deal with the potential formation of shocks, the authors
made use of high resolution shock capturing methods.
In that reference, the authors found certain functional forms
of the theory that appears to allow 
for evolution without the formation
of elliptic regions, even in the strongly coupled regime.
\subsection{Cubic Horndeski gravity
\label{sec:cubic_horndeski_numerical}}
Figueras and Fran\c{c}a\cite{Figueras:2020dzx,Figueras:2021abd}
have recently performed the first fully-nonlinear simulations
of dynamical black hole spacetimes in a cubic Horndeski gravity theory.
In both works, they considered couplings of the form
$\mathcal{G}_2=g_2X^2$ and $\mathcal{G}_3=g_3X$, where $g_2$ and
$g_3$ are constants, and either $V=0$\cite{Figueras:2020dzx}
or $V=\frac{1}{2}m^2\phi^2$\cite{Figueras:2021abd}. 
To evolve the equations of motion in well-posed fashion,
the authors rewrote the equations of motion as
described by Kovacs\cite{Kovacs:2019jqj}, and then made
use of the CCZ4 formulation\cite{Bona:2003fj,Alic:2011gg} 
(see also Sec.~\ref{sec:cubic_horndeksi_gravity_well_posedness}).

From spherical collapse simulations  
Figueras and Fran\c{c}a found that the equations of motion
for cubic Horndeski gravity can dynamically form
a sonic line and elliptic region outside of
a black hole horizon\cite{Figueras:2020dzx}.
This loss of hyperbolicity is located in the scalar sector of the theory.
Despite the loss of hyperbolicity in the strongly coupled regime,
the theory did remain hyperbolic in regions where the solution
remained safely in the weakly-coupled regime.
As the authors used CCZ4 puncture-type evolution\cite{Andrade:2021rbd}, 
to avoid hyperbolicity
issues in the interior of the black hole they smoothly turned
off the coefficients $g_2$ and $g_3$ inside the apparent horizon.
In particular, they multiplied the coefficients by a sigmoid function
\begin{align}
   g_i
   \to
   g_i
   \times 
   \frac{1}{1 + e^{-\frac{2}{w}\left(\left(x/\bar{x}\right)-1\right)}}
   ,
\end{align}
where $x$ is a distance measure from the ``center'' of the black hole
to the apparent horizon, 
and $\bar{x},w$ are pre-set constants\footnote{While the sigmoid will
not be exactly one for all $x<\infty$, it rapidly transitions
to values very close to $1$ for $x$ outside of a narrow
``transition region'' whose location and width are
determined by $\bar{x}$ and $w$, respectively.}.
As a measure of the distance inside the black hole,
Figueras and Fran\c{c}a used the value of the conformal
factor $\chi$ in the CCZ4 formulation, the level sets of
which have been empirically shown to closely follow the
apparent horizons of black hole and black 
strings\cite{Bantilan:2019bvf,Andrade:2020dgc}
(for example, for Schwarzschild black holes in moving
puncture gauge, within the CCZ4 formulation the
black hole apparent horizon settles to around 
$\chi\sim0.25$\cite{Figueras:2020dzx}).

Following this work, Figueras and Fran\c{c}a considered
binary black hole collisions for the same kinds of 
theories\cite{Figueras:2021abd}.
In this work, initial data was set by super-imposing two boosted 
spherically symmetric solutions of lumps of scalar field $\phi$. 
While this does not exactly
solve the constraints, provided the black holes are far enough
apart, it provides a reasonably close solution to the full
equations of motion.
The two lumps of scalar field quickly collapsed to forming black holes
with some leftover scalar field energy outside of the black holes.
While black holes cannot support scalar field hair for the class of
Horndeski theories they consider\cite{Hui:2012qt,Maselli:2015yva}, 
the initial data the authors choose allowed
for some residual scalar field during the inspiral, which they showed 
lead to some dephasing of eccentric black hole binaries
as compared to GR\cite{Figueras:2021abd}.
\subsection{$4\partial ST$ gravity (alias
Einstein scalar Gauss-Bonnet (ESGB) gravity\label{sec:esgb_numerical})}
Einstein scalar Gauss-Bonnet (ESGB) gravity
(See Eq.~\eqref{eq:action_4dST_gravity}) 
has received much attention recently as variants of
ESGB gravity allow for scalar ``hairy'' black 
holes\cite{Kanti:1995vq,Sotiriou:2013qea,Sotiriou:2014pfa}.
This can be seen by considering the scalar field equation of motion
(Eq.~\eqref{eq:eom_4dst_scalar})
\begin{align}
   \label{eq:scalar_eom_4est_again}
   E^{(\phi)}
   &=
   \Box\phi - V'
   \nonumber\\
   &+
   3\alpha X\Box\phi
   -
   2\alpha 
   \nabla^{\alpha}\phi\nabla^{\beta}\phi\nabla_{\alpha}\nabla_{\beta}\phi
   -
   3\alpha' X^2
   \nonumber\\
   &+
   \beta'\mathcal{R}_{GB}
   .
\end{align}
For Schwarzschild and Kerr black holes, $\mathcal{R}_{GB}\neq0$,
thus even if $\phi=\partial_0\phi=0$ at $t=0$, provided
$\beta'\left(0\right)\neq0$, the scalar field sees a nonzero
``source'' term, which causes it to grow about the black hole.
For example, in ``shift-symmetric scalar Gauss-Bonnet gravity''
$V=\alpha=0$ and $\beta=\lambda\phi$, where $\lambda$ is
a constant. This theory has scalar-hairy black holes,
while neutron stars in the theory remain relatively less
hairy, so binary black hole systems may provide the strongest
constraints on these theories\cite{Yagi:2015oca}.
This theory has been extensively studied,
using the order reduction approach\cite{Benkel:2016rlz,Benkel:2016kcq,
Witek:2018dmd,Okounkova:2019zep,Okounkova:2020rqw},
in the static limit\cite{Sotiriou:2014pfa,Sullivan:2020zpf}, 
and restricted to spherically
symmetric spacetimes\cite{Ripley:2019irj,Ripley:2019aqj}.
The first fully nonlinear simulations of this theory of binary
black hole spacetimes were completed using the MGH 
formulation, with vacuum GR black hole initial data\cite{East:2020hgw}.

Even if $\beta'\left(0\right)=0$, GR black holes can still
be unstable to 
\emph{spontaneous black hole scalarization}\cite{Doneva:2017bvd,
Silva:2017uqg,
Minamitsuji:2018xde,Silva:2018qhn,
Dima:2020yac,Herdeiro:2020wei,Berti:2020kgk}
(which is analogous to spontaneous scalarization for
neutron stars in scalar-tensor gravity\cite{Damour:1996ke}--see also 
Sec.~\ref{sec:scalar_tensor_special_case_numerical}).
This can be seen by expanding Eq.~\eqref{eq:scalar_eom_4est_again}
in powers of $\phi$ (assuming $\phi\ll1$):
\begin{align}
   0
   =
   \Box\phi 
   -
   \left(V'' - \beta''\mathcal{R}_{GB}\right)\phi
   +
   \mathcal{O}\left(\phi^2\right)
   .
\end{align}
We see that $\beta''\mathcal{R}_{GB}$ can act as an ``effective''
mass, that could lead to a tachyonic instability depending on
the sign of that term.
There have been several recent numerical simulations of binary
black hole systems that exhibit spontaneous black hole scalarization
(and de-scalarization), using order reduction 
methods\cite{Silva:2020omi,Elley:2022ept}\footnote{We note that
one potential limitation of the perturbative approach for theories
that are dynamically unstable to scalarization is that nonlinear
effects are generally needed to \emph{saturate} the instability.}.
Fully nonlinear simulations of binary black holes
in a variant of this theory that exhibits black hole scalarization
has only been performed for head-on collisions\cite{East:2021bqk}.

Evolving exact solutions
through the merger for scalar-hairy black holes in this
theory (that is when $\beta\neq0$) has proven to be challenging, 
as during the merger phase, spacetime curvature gradients can grow
large, which could trigger the formation of new elliptic regions.
Provided these new elliptic regions remain inside the event horizon
of the final black hole a given theory can remain theoretically viable,
but excising these regions using black hole excision can be challenging
before an apparent horizon for the final black hole has been found.

The study of binary black hole systems for other values of scalar
field potential $V$ and coupling $\alpha$ remains relatively
less studied, at least together with nontrivial values of $\beta$. 

\section{Final remarks\label{sec:final_remarks}}

We have reviewed the status of numerically constructing 
black hole/neutron star solutions in Horndeski theories of gravity.
We discussed exact and perturbative
methods that give well-posed formulations of the evolution and constraint
equations, at least for so-called \emph{weakly-coupled} solutions.
Both the exact and perturbative approaches introduce new difficulties
not encountered in numerical solutions of the Einstein equations. 
Exact solutions may suffer from the formation of shocks or
a loss of hyperbolicity, while perturbative solutions must contend
with the secular growth of errors in time.
Many basic questions about the Horndeski theories remain unanswered,
especially in the context of binary black hole and neutron star mergers,
and much work remains to properly understand how to (numerically)
solve and interpret the equations of motion for these theories.
Nevertheless, there are several promising ways to extend numerical relativity  
methods to construct strong gravity, weakly coupled 
solutions to the Horndeski gravity theories.

\section*{Acknowledgments}

 I thank  
 Aron Kovacs, Frans Pretorius,
 Harvey Reall, Leo Stein, Maxence Corman, 
 Nicolas Yunes, Ulrich Sperhake,  and William East
 for helpful discussions,
 and thank Abhishek Hegde, Aron Kovacs, 
 Tomas Galvez-Ghersi, Maxence Corman, 
 Tiago Fran\c{c}a, and William East
 for helpful comments on an earlier
 draft of this article.
 While writing this article I was
 supported by STFC Research Grant No. ST/V005669/1.

\appendix

\section{Equations of motion for Horndeski gravity
   \label{sec:eom_horndeski_gravity}
}
Varying \eqref{eq:general_horndeski_action} 
with respect to the metric and scalar fields, 
we obtain the tensor and scalar equations of motion,
respectively. These were first derived by Horndeski\cite{horndeski_paper}.
We have checked the equations in Papallo\cite{Papallo:2017ddx}
(See also Tanahashi and Ohashi\cite{Tanahashi:2017kgn}
for the ``shift-symmetric'' Horndeski equations of motion, which are
when the $\mathcal{G}_i$ only depend on derivatives of $\phi$)
using the \texttt{xTensor}\cite{xAct,martin2008xperm,Brizuela:2008ra} 
package for \texttt{Mathematica}.
Due to our different normalization for $R$, there is a
factor of $2$ difference between the tensor equations of motion
presented here and those presented in Papallo\cite{Papallo:2017ddx}.
Our Mathematica notebook, along with other code used to generate
some of the plots used in this article,
can be accessed online\cite{github_article}.
\begin{align}
\label{eq:tensor_eom_horndeski}
   \left(E^{(g)}\right)^{\alpha}{}_{\beta}
   &=
   \nonumber\\
   -&
   \frac{1}{4}\left(
      1 
      + 
      2\mathcal{G}_4 
      - 
      4X\partial_X\mathcal{G}_4 
      + 
      2X \partial_{\phi}\mathcal{G}_5
   \right)
   \delta^{\alpha\gamma_1\gamma_2}_{\beta\delta_1\delta_2}
   R_{\gamma_1\gamma_2}{}^{\delta_1\delta_2}
   \nonumber\\
   +&
   \frac{1}{2}\left(
      \partial_X\mathcal{G}_4
      -
      \partial_{\phi}\mathcal{G}_5
   \right)
   \delta^{\alpha\gamma_1\gamma_2\gamma_3}_{\beta\delta_1\delta_2\delta_3}
   \nabla_{\gamma_1}\phi\nabla^{\delta_1}\phi
   R_{\gamma_2\gamma_3}{}^{\delta_2\delta_3}
   \nonumber\\
   -&
   \frac{1}{2}
   \left(X\partial_X\mathcal{G}_5\right)
   \delta^{\alpha\gamma_1\gamma_2\gamma_3}_{\beta\delta_1\delta_2\delta_3}
   \nabla_{\gamma_1}\nabla^{\delta_1}\phi
   R_{\gamma_2\gamma_3}{}^{\delta_2\delta_3}
   \nonumber\\
   -&
   \left(
      -
      V
      +
      X
      +
      \mathcal{G}_2
      +
      2X\partial_{\phi}\mathcal{G}_3
      +
      4X\partial_{\phi}^2\mathcal{G}_4
   \right)
   \delta^{\alpha}_{\beta}
   \nonumber\\
   -&
   \left(
      1
      +
      \partial_X\mathcal{G}_2
      +
      2\partial_{\phi}\mathcal{G}_3
      +
      2\partial_{\phi}^2\mathcal{G}_4
   \right)
   \nabla_{\beta}\phi\nabla^{\alpha}\phi
   \nonumber\\
   +&
   2\left(
      X\partial_X\mathcal{G}_3
      +
      \partial_{\phi}\mathcal{G}_4
      +
      2X\partial_X\partial_{\phi}\mathcal{G}_4
   \right)
   \delta^{\alpha\gamma}_{\beta\delta}
   \nabla_{\gamma}\nabla^{\delta}\phi
   \nonumber\\
   +&
   \left(
      \partial_X\mathcal{G}_3
      +
      4\partial_X\partial_{\phi}\mathcal{G}_4
      -
      \partial_{\phi}^2\mathcal{G}_5
   \right)
   \delta^{\alpha\gamma_1\gamma_2}_{\beta\delta_1\delta_2}
   \nabla_{\gamma_1}\phi\nabla^{\delta_1}\phi
   \nabla_{\gamma_2}\nabla^{\delta_2}\phi
   \nonumber\\
   +&
   \left(
      \partial_X\mathcal{G}_4
      +
      2X\partial_X^2\mathcal{G}_4
      -
      \partial_{\phi}\mathcal{G}_5
      -
      X\partial_X\partial_{\phi}\mathcal{G}_5
   \right)
   \delta^{\alpha\gamma_1\gamma_2}_{\beta\delta_1\delta_2}
   \nabla_{\gamma_1}\nabla^{\delta_1}\phi
   \nabla_{\gamma_2}\nabla^{\delta_2}\phi
   \nonumber\\
   +&
   \left(
      \partial_X^2\mathcal{G}_4
      -
      \partial_X\partial_{\phi}\mathcal{G}_5
   \right)
   \delta^{\alpha\gamma_1\gamma_2\gamma_3}_{\beta\delta_1\delta_2\delta_3}
   \nabla_{\gamma_1}\nabla^{\delta_1}\phi
   \nabla_{\gamma_2}\nabla^{\delta_2}\phi
   \nabla_{\gamma_3}\phi\nabla^{\delta_3}\phi
   \nonumber\\
   -&
   \frac{1}{3}
   \left(
      \partial_X\mathcal{G}_5
      +
      X\partial_X^2\mathcal{G}_5
   \right)
   \delta^{\alpha\gamma_1\gamma_2\gamma_3}_{\beta\delta_1\delta_2\delta_3}
   \nabla_{\gamma_1}\nabla^{\delta_1}\phi
   \nabla_{\gamma_2}\nabla^{\delta_2}\phi
   \nabla_{\gamma_3}\nabla^{\delta_3}\phi
   ,\\
\label{eq:scalar_eom_horndeski}
   E^{(\phi)}
   &=
   \nonumber\\
   -&
   \left(
      1
      +
      \partial_X\mathcal{G}_2
      +
      2X\partial_X^2\mathcal{G}_2
      +
      2\partial_{\phi}\mathcal{G}_3
      +
      2X\partial_X\partial_{\phi}\mathcal{G}_3
   \right)\Box\phi
   \nonumber\\
   -&
   \left(
      \partial_X^2\mathcal{G}_2
      +
      2\partial_X\partial_{\phi}\mathcal{G}_3
      +
      2\partial_X\partial_{\phi}^2\mathcal{G}_4
   \right)
   \delta^{\gamma_1\gamma_2}_{\delta_1\delta_2}
   \nabla_{\gamma_1}\phi\nabla^{\delta_1}\phi
   \nabla_{\gamma_2}\nabla^{\delta_2}\phi
   \nonumber\\
   -&
   \left(
      \partial_X\mathcal{G}_3
      +
      X\partial_X^2\mathcal{G}_3
      +
      2X\partial_X^2\partial_{\phi}\mathcal{G}_4
      +
      3\partial_X\partial_{\phi}\mathcal{G}_4
   \right)
   \delta^{\gamma_1\gamma_2}_{\delta_1\delta_2}
   \nabla_{\gamma_1}\nabla^{\delta_1}\phi
   \nabla_{\gamma_2}\nabla^{\delta_2}\phi
   \nonumber\\
   -&
   \frac{1}{4}\left(
      \partial_X\mathcal{G}_3
      +
      4\partial_X\partial_{\phi}\mathcal{G}_4
      -
      \partial_{\phi}^2\mathcal{G}_5
   \right)
   \delta^{\gamma_1\gamma_2\gamma_3}_{\delta_1\delta_2\delta_3}
   \nabla_{\gamma_1}\phi\nabla^{\delta_1}\phi
   R_{\gamma_1\gamma_2}{}^{\delta_1\delta_2}
   \nonumber\\
   -&
   \left(
      X\partial_X\mathcal{G}_3
      +
      \partial_{\phi}\mathcal{G}_4
      +
      2X\partial_X\partial_{\phi}\mathcal{G}_4
   \right)R
   \nonumber\\
   -&
   \frac{1}{2}\left(
      \partial_X^2\mathcal{G}_3
      +
      4\partial_X^2\partial_{\phi}\mathcal{G}_4
      -
      \partial_X\partial_{\phi}^2\mathcal{G}_5
   \right)
   \delta^{\gamma_1\gamma_2\gamma_3}_{\delta_1\delta_2\delta_3}
   \nabla_{\gamma_1}\nabla^{\delta_1}\phi
   \nabla_{\gamma_2}\nabla^{\delta_2}\phi
   \nabla_{\gamma_3}\phi\nabla^{\delta_3}\phi
   \nonumber\\
   -&
   \frac{1}{2}\left(
      \partial_X\mathcal{G}_4
      +
      2X\partial_X^2\mathcal{G}_4
      -
      \partial_{\phi}\mathcal{G}_5
      -
      X\partial_X\partial_{\phi}\mathcal{G}_5
   \right)
   \delta^{\gamma_1\gamma_2\gamma_3}_{\delta_1\delta_2\delta_3}
   \nabla_{\gamma_1}\nabla^{\delta_1}\phi
   R_{\gamma_2\gamma_3}{}^{\delta_2\delta_3}
   \nonumber\\
   -&
   \frac{1}{2}\left(
      \partial_X^2\mathcal{G}_4
      -
      \partial_X\partial_{\phi}\mathcal{G}_5
   \right)
   \delta^{\gamma_1\gamma_2\gamma_3\gamma_4}_{\delta_1\delta_2\delta_3\delta_4}
   \nabla_{\gamma_1}\nabla^{\delta_1}\phi
   \nabla_{\gamma_2}\phi\nabla^{\delta_2}\phi
   R_{\gamma_3\gamma_4}{}^{\delta_3\delta_4}
   \nonumber\\
   -&
   \frac{1}{3}\left(
      3\partial_X^2\mathcal{G}_4
      +
      2X\partial_X^3\mathcal{G}_4
      -
      2\partial_x\partial_{\phi}\mathcal{G}_5
      -
      X\partial_X^2\partial_{\phi}\mathcal{G}_5
   \right)
   \delta^{\gamma_1\gamma_2\gamma_3}_{\delta_1\delta_2\delta_3}
   \nabla_{\gamma_1}\nabla^{\delta_1}\phi
   \nabla_{\gamma_2}\nabla^{\delta_2}\phi
   \nabla_{\gamma_3}\nabla^{\delta_3}\phi
   \nonumber\\
   -&
   \frac{1}{3}\left(
      \partial_X^3\mathcal{G}_4
      -
      \partial_X^2\partial_{\phi}\mathcal{G}_5
   \right)
   \delta^{\gamma_1\gamma_2\gamma_3\gamma_4}_{\delta_1\delta_2\delta_3\delta_4}
   \nabla_{\gamma_1}\nabla^{\delta_1}\phi
   \nabla_{\gamma_2}\nabla^{\delta_2}\phi
   \nabla_{\gamma_3}\nabla^{\delta_3}\phi
   \nabla_{\gamma_4}\phi\nabla^{\delta_4}\phi
   \nonumber\\
   +&
   \frac{1}{12}\left(
      2\partial_X^2\mathcal{G}_5
      +
      X\partial_X^3\mathcal{G}_5
   \right)
   \delta^{\gamma_1\gamma_2\gamma_3\gamma_4}_{\delta_1\delta_2\delta_3\delta_4}
   \nabla_{\gamma_1}\nabla^{\delta_1}\phi
   \nabla_{\gamma_2}\nabla^{\delta_2}\phi
   \nabla_{\gamma_3}\nabla^{\delta_3}\phi
   \nabla_{\gamma_4}\nabla^{\delta_4}\phi
   \nonumber\\
   +&
   \frac{1}{4}\left(
      \partial_X\mathcal{G}_5
      +
      X\partial_X^2\mathcal{G}_5
   \right)
   \delta^{\gamma_1\gamma_2\gamma_3\gamma_4}_{\delta_1\delta_2\delta_3\delta_4}
   \nabla_{\gamma_1}\nabla^{\delta_1}\phi
   \nabla_{\gamma_2}\nabla^{\delta_2}\phi
   R_{\gamma_3\gamma_4}{}^{\delta_3\delta_4}
   \nonumber\\
   +&
   \frac{1}{16}\left(X\partial_X\mathcal{G}_5\right)
   \delta^{\gamma_1\gamma_2\gamma_3\gamma_4}_{\delta_1\delta_2\delta_3\delta_4}
   R_{\gamma_1\gamma_2}{}^{\delta_1\delta_2}
   R_{\gamma_3\gamma_4}{}^{\delta_3\delta_4}
   \nonumber\\
   +&
   2X\left(
      \partial_{\phi}^2\mathcal{G}_3
      +
      \partial_X\partial_{\phi}\mathcal{G}_2
   \right)
   X
   -
   \partial_{\phi}\mathcal{G}_2
   +
   \partial_{\phi}V
   .
\end{align}
While $4\partial ST$ gravity can be written in terms of a Horndeski
theory, for completeness we list its equations of motion here as well
\begin{align}
   \label{eq:eom_4dst_tensor}
   E^{\alpha}{}_{\beta}
   &=
   R^{\alpha}{}_{\beta}
   -
   \frac{1}{2}\delta^{\alpha}_{\beta}R
   -
   \nabla^{\alpha}\phi\nabla_{\beta}\phi
   +
   \left(-X+V\right)\delta^{\alpha}_{\beta}
   \nonumber\\
   &-
   2\alpha X\nabla^{\alpha}\phi\nabla_{\beta}\phi
   -
   \alpha X^2\delta^{\alpha}_{\beta}
   \nonumber\\
   &+
   2\delta^{\mu\rho\gamma\alpha}_{\nu\sigma\delta\beta}
   R^{\nu\sigma}{}_{\mu\rho}
   \nabla^{\delta}\nabla_{\gamma}\beta
   ,\\
   \label{eq:eom_4dst_scalar}
   E^{(\phi)}
   &=
   \Box\phi - V'
   \nonumber\\
   &+
   3\alpha X\Box\phi
   -
   2\alpha 
   \nabla^{\alpha}\phi\nabla^{\beta}\phi\nabla_{\alpha}\nabla_{\beta}\phi
   -
   3\alpha' X^2
   \nonumber\\
   &+
   \beta'\mathcal{R}_{GB}
   .
\end{align}

\section{Hyperbolicity and well-posedness of evolution (hyperbolic) 
partial differential equations
\label{sec:review_hyperbolicity}
}

For completeness, here we review some general concepts about the
initial value problem for hyperbolic systems of partial
differential equations (PDEs),
several different notions of hyperbolicity, and state
several theorems that relate hyperbolicity and the well-posedness
of the initial value problem.
There are many reviews on this subject, some of which are
specialized for application in mathematical/numerical 
relativity\cite{courant1962methods,whitham2011linear,
kreiss_lorenz,Sarbach:2012pr,Hilditch:2013sba}, so we
will only briefly outline the main ideas.

We consider systems of first order PDEs that take the form
\begin{equation}
\label{eq:system_first_order_pde}
   A^I_J\left(x^{\alpha},v^K,\partial_iv\right)\partial_0v^J
   +
   B^I\left(x^{\alpha},v^K,\partial_iv^L\right)
   =
   0
   ,
\end{equation} 
where $I,J,..=1,...,N$ index the $N$ 
equations of motion and dynamical fields $v^J$,
$i,j,...=1,...,n$ index the $n$ spatial coordinates, 
and $\alpha,\beta,...=0,...,n$ index
the $n+1$ spacetime coordinates.
We note that through field redefinitions, we can write
systems of second order equations that are linear in $\partial_0^2v^J$
\begin{align}
\label{eq:second_order_general}
   Y^I_J\left(
      x^{\alpha},
      v^K,
      \partial_0v^L,
      \partial_iv^M,
      \partial_i\partial_0v^N,
      \partial_i\partial_jv^P
   \right)
   \partial_0^2v^J
   &\nonumber\\
   +
   Z^I\left(
      x^{\alpha},
      v^K,
      \partial_0v^L,
      \partial_iv^M,
      \partial_i\partial_0v^N,
      \partial_i\partial_jv^P
   \right)
   &=
   0
   ,
\end{align}
in the form 
\eqref{eq:system_first_order_pde}
\cite{Sarbach:2012pr,Hilditch:2013sba,Kovacs:2020pns}.
The equations of motion for Einstein gravity, 
Horndeski gravity, Lovelock
gravity, and Einstein-Maxwell theory all can be
written in the form of
\eqref{eq:second_order_general}
\cite{Papallo:2017qvl,Kovacs:2020pns,Davies:2021frz}.

Provided $A^I_J$ is invertible, we can rewrite
Eq.~\eqref{eq:system_first_order_pde} as
\begin{align}
\label{eq:system_first_order_pde_C}
   \partial_0v^I
   +
   C^I\left(x^{\alpha},v^J,\partial_iv^K\right)
   =
   0
   .
\end{align}
where
\begin{align}
   C^I_J
   \equiv
   \left(A^{-1}\right)^I_KB^K
   .
\end{align}
For the remainder of this section we will assume this to be the case.

\subsection{Well-posedness
\label{sec:well_posedness}
}

We say that the system \eqref{eq:system_first_order_pde} has a
\emph{well-posed initial value problem (well posed IVP)} if there
exist constants $A$ and $\alpha$ such that for all smooth enough
initial data $f^I$, the solution remains uniformly bounded by an exponential
\begin{align}
   \left|v^I\right|(t)
   \leq
   A e^{\alpha t} \left|f^I\right|
   .
\end{align}
Here we use the $L^2$ norm
\begin{align}
   \left|v^I\right|(t)
   \equiv
   \sqrt{
      \int_{\mathbb{R}^3} d^3x \left(v_I\right)^{\dagger} v^I
   }
   .
\end{align}
This condition guarantees that the
solution depends continuously on the initial data. 
\subsection{Characteristics and notions of hyperbolicity
\label{sec:characteristics_hyperbolicity}
}
The \emph{principal symbol} is defined to be
\begin{equation}
   \mathcal{P}^I_J\left(\xi_{\alpha}\right)
   \equiv 
   \delta^I_J\xi_0
   +
   \left(
      \frac{
         \delta C^I
      }{
         \delta(\partial_iv^J)
      }
   \right)
   \xi_i
   ,
\end{equation}
where $\xi_{\alpha}$ is an $n$ dimensional unit covector,
and $\xi_i$ is a unit spatial covector.
A \emph{characteristic
surface} $\Sigma\subset M$ is spanned by
co-vectors that satisfy the \emph{characteristic equation}
\begin{equation}
\label{eq:characteristic_equation_chptr_2}
   \mathrm{det}\left(\mathcal{P}^I_J\left(\xi_{\alpha}\right)\right)
   = 
   0
   .
\end{equation}
   Replacing $\xi_a$ with $\partial_a$, one
obtains from \eqref{eq:characteristic_equation_chptr_2}
the \emph{eikonal equation} 
for the characteristic surface\cite{christodoulou2008mathematical}. 
Finally, for the $n^{th}$ real solution to the characteristic equation,
we define the \emph{characteristic velocity} to be
\begin{align}
   c_i^{(n)} 
   \equiv
   -
   \frac{\xi_i^{(n)}}{\xi_0^{(n)}}
   .
\end{align}
We define the \emph{characteristic speed} to be the 2-norm of the
characteristic velocity, $c\equiv \sqrt{\sum_i |c_i|^2}$.

In order to determine the \emph{hyperbolicity} of the
system \eqref{eq:system_first_order_pde_C}, we linearize the equations
about a background solution $v_0^I$, about a given point 
$x^{\alpha}=x_0^{\alpha}$:
\begin{align}
   v^I
   \to&
   v^I_0 + \epsilon u^I
   ,
\end{align}
where $\epsilon\ll1$.
We then have a linear PDE for $u^I$: 
\begin{align}
\label{eq:general_linearized_equation}
   \left(
      \delta^I_J
      \partial_0
      +
      \left[F^I_J\right]^i\partial_i
   \right)
   u^J
   +
   G^I_Ju^J
   =
   0
   ,
\end{align}
where
\begin{align}
   \left[F^I_J\right]^i
   \equiv
   \frac{\delta C^I}{\delta \left(\partial_iv^J\right)}\Bigg|_{v^I=v^I_0}
   ,\qquad
   G^I_J
   \equiv
   \frac{\delta C^I}{\delta v^J}\Bigg|_{v^I=v^I_0}
   .
\end{align}

We now define different notions of hyperbolicity.
Provided $A^I_J$ is invertible, at $x^{\alpha}=x^{\alpha}_0$
and $v^I=v^I_0$ the system \eqref{eq:system_first_order_pde}
is \emph{weakly hyperbolic} if all the eigenvalues
of $\left[F^I_J\right]^i\xi_i$ are real for all unit $\xi_i$.
The system is \emph{strongly hyperbolic} if $\left[F^I_J\right]^i\xi$
has a complete set of eigenvectors that have real eigenvalues,
and that the matrix
whose columns are the eigenvectors, $T^I_J\left(\xi_i\right)$, satisfies
\begin{align}
\label{eq:strong_hyperbolicity_eigenvectors}
   \left| T^I_J\left(\xi_i\right)\right|
   +
   \left| \left(T^{-1}\right)^I_J\left(\xi_i\right)\right|
   \leq
   \Lambda
   ,
\end{align}
where $\Lambda$ is a constant independent of $\xi_i$.
Here $\left|\cdots\right|$ is a matrix norm.
If the eigenvectors depend continuously on $\xi_i$,
then \eqref{eq:strong_hyperbolicity_eigenvectors} is satisfied
since $\xi_i$ is varied over a compact set
(recall we impose that $\xi_i$ has unit norm).
Sometimes the eigenvectors are called 
``eigenfields''\cite{alcubierre2008introduction}.

Before continuing, we mention another (equivalent) definition of strong
hyperbolicity that is commonly used 
(see \refcite{kreiss_lorenz,Papallo:2017qvl,Kovacs:2020pns,Kovacs:2021vdk}). 
The system \eqref{eq:system_first_order_pde} is strongly hyperbolic if
there is a smooth, positive definite $N\times N$
Hermitian matrix $H^I_J(\xi_i)$--the \emph{(Kreiss-)symmetrizer}--and
a positive constant $\Lambda$ such that for all $\xi_i$
\begin{align}
   H^I_K(\xi_j) \left[F^K_J\right]^i\xi_i
   =
   \left(\left[F^I_K\right]^i\xi_i\right)^{\dagger}  H^K_J (\xi_j) 
   ,
\end{align}
and
\begin{align}
   \frac{1}{\Lambda} \left|\delta^I_J\right| 
   \leq 
   \left|H^I_J(\xi_j)\right| 
   \leq  
   \Lambda \left|\delta^I_J\right|
   .
\end{align}
This definition is equivalent to the earlier definition
as one can show that
\begin{align}
   H^I_J 
   \equiv 
   \left(\left(V^{-1}\right)^{\dagger}\right)^I_K
   \left(V^{-1}\right)^K_J, 
\end{align}
where the columns of $V^I_J$ are
the eigenvectors to $\left[F^I_J\right]^i\xi_i$, is a symmetrizer.

We can relate strong hyperbolicity to
well-posedness of the initial value problem for systems
of the form \eqref{eq:system_first_order_pde_C} with the following theorem
(for a proof, see Chapter 5 of Taylor\cite{Taylor1991}; note
that in that reference the author calls strong hyperbolicity
\emph{symmetrizable hyperbolicity}):
\begin{enumerate}
   \item [] \textbf{Theorem}
   For strongly hyperbolic systems \eqref{eq:system_first_order_pde_C}, 
   the Cauchy problem with initial data 
   $v^I (0, \xi) = f^I (\xi)$ is well-posed in Sobolev
   spaces $\mathbb{H}^s$ with $s > s_0$ for some constant $s_0$. 
   That is to say, there exists a unique local solution
   $v^I \in C([0, T ), \mathbb{R}^d)$ with $T > 0$ 
   depending on the $\mathbb{H}^s$-norm of the initial data.
\end{enumerate}
The Sobolev space $\mathbb{H}^s$ is the space of functions that
have a finite Sobolev norm
\begin{align}
   \left|v\right|_{2,s}
   \equiv
   \sqrt{
      \sum_{k=0}^s
      \int_{\mathbb{R}^3}d^3x 
      \left|\partial_i^{K}v\right|^2
   }
   ,
\end{align}
where $\partial_i^{K}$ schematically stands for the sum over all
spatial derivatives of order $k$. 

In order words, for sufficiently regular initial data, a strongly hyperbolic
system of PDE has a well-posed IVP.
In physics applications the initial data is often very smooth (unless
the initial data has ``shocks''),
so we will not discuss issues about optimal regularity of initial data.

\subsection{Physical interpretation of characteristics
\label{sec:physical_interpretation_characteristics}
}

We end by briefly reviewing a ``physical'' interpretation for
characteristic surfaces.
First we rewrite \eqref{eq:general_linearized_equation} as
\begin{align}
   \mathcal{P}^I_J\left(\partial_{\alpha}\right)u^J
   +
   G^I_Ju^J
   =
   0
   .
\end{align}
We next consider high frequency plane wave-like solutions:
\begin{align}
   u^I(x^{\alpha})
   =
   \tilde{u}^I(x^i)e^{-ik_{\alpha}x^{\alpha}/\epsilon}
   ,
\end{align}
where $\epsilon\ll1$.
To leading order in $\epsilon$ we see that 
\begin{align}
   \mathcal{P}^I_J\left(k_{\alpha}\right)u_0^J
   =
   \left(
      \delta^I_Jk_t
      +
      \left[F^I_J\right]^ik_i
   \right)
   \tilde{u}^J
   =
   0
   .
\end{align}
Nontrivial solutions to this equation exist only if
$\det\mathcal{P}^I_J\left(k_{\alpha}\right)=0$, that is only
if the wave vector $k_{\alpha}$ is a characteristic vector.
The corresponding functional solutions $\tilde{u}^I$ are
the eigenvectors to $\left[F^I_J\right]^i\xi_i$, 
and are advected along the characteristics.
Thus the wave fronts of high-frequency small amplitude solutions
propagate on the characteristic surfaces.
We also see that
any low-amplitude, high frequency solution 
can be locally be written as a linear
combination of the eigenvectors, 
provided the system is strongly hyperbolic.

\section{Mixed-type partial differential equations
   \label{sec:mixed_type}
}

For completeness, here we review \emph{mixed-type} PDE. 
Mixed-type PDE are PDE which have degrees of freedom which
can change character from being hyperbolic to being elliptic
in different solution regions;
for general reviews see
\refcite{rassias1990lecture,otway2015elliptic,doi:10.1142/S0219891604000081}.

\subsection{Two-dimensional PDE
   \label{sec:two_dimensional_mixed}
}
\emph{Mixed-type} PDE are differential equations
that are neither hyperbolic nor elliptic over their 
entire solution domain\cite{rassias1990lecture,otway2015elliptic}.
The simplest examples of mixed-type PDE arise in two dimensions,
and  much that is formally known about these theories appear in this context.
For these reasons we will focus mostly on two-dimensional mixed-type equations.

We first review some terminology\cite{evans2010partial}.
Consider PDE of the form
\begin{align}
\label{eq:simple_2d}
   \mathcal{A}\left(x,y,u,\partial_xu,\partial_yu\right)\partial_x^2u(x,y)
   +
   \mathcal{B}\left(x,y,u,\partial_xu,\partial_yu\right)\partial_x\partial_yu(x,y)
   &
   \nonumber\\
   +
   \mathcal{C}\left(x,y,u,\partial_xu,\partial_yu\right)\partial_y^2u(x,y)
   +
   \mathcal{F}\left(x,y,u,\partial_xu,\partial_yu\right)
   &=
   0
   .
\end{align}
The characteristic equation is
\begin{align}
   \mathcal{A}\xi_x^2
   +
   \mathcal{B}\xi_x\xi_y
   +
   \mathcal{C}\xi_y^2
   =
   0
   .
\end{align}
Following the discussion in \ref{sec:review_hyperbolicity}, we see that
the theory is hyperbolic provided we can find two real solutions to this
equation.
This in turn is dictated by the \emph{discriminant}, which is defined to be
\begin{align}
   \mathcal{D}
   \equiv
   \mathcal{B}^2
   -
   4\mathcal{A}\mathcal{C}
   .
\end{align}
Consider a point $(x,y)$, and say we have specified
$u,\partial_xu,\partial_yu$ at that point.
We say that Eq.~\eqref{eq:simple_2d} is
\emph{hyperbolic} if $\mathcal{D}>0$, \emph{parabolic} if $\mathcal{D}=0$,
and \emph{elliptic} if $\mathcal{D}<0$.
The classic examples of hyperbolic, parabolic, and elliptic PDEs in two
dimensions are
\begin{subequations}
\label{eq:canonical_form}
\begin{align}
   \partial_x^2u - \partial_y^2u
   &=
   0
   ,\\
   \partial_xu - \partial_y^2u
   &=
   0
   ,\\
   \partial_x^2u + \partial_y^2u
   &=
   0
   .
\end{align}
\end{subequations}
Locally we can transform any hyperbolic, parabolic, and elliptic
equation to take the forms \eqref{eq:canonical_form}
through coordinate transformations.

The canonical examples of mixed-type PDE are the Tricomi
and Keldysh equations\cite{chen2015tricomi}, which respectively are 
\begin{align}
   \partial_x^2u(x,y)
   +
   x
   \partial_y^2u(x,y)
   &=
   0
   ,\\
   \partial_x^2u(x,y)
   +
   \frac{1}{x}
   \partial_y^2u(x,y)
   &=
   0
   .
\end{align}
These equations are hyperbolic/parabolic/elliptic when 
$x<0 \; / \; x=0 \; / \; x > 0$.
The hyperbolic and elliptic regions are separated by the parabolic
\emph{sonic line}\footnote{This terminology comes from hydrodynamics,
which is the physical context in which mixed-type PDE were
first studied\cite{bams/1183548680,doi:10.1142/S0219891604000081}.
See also \refcite{doi:10.2514/3.6131} for a discussion of numerical
methods to solve mixed-type equations as they appear in
hydrodynamics.}.
Locally we can transform mixed-type PDE near their sonic lines
into either a Tricomi or Keldysh form.

\begin{figure}[h]
\includegraphics[width=0.45\textwidth]{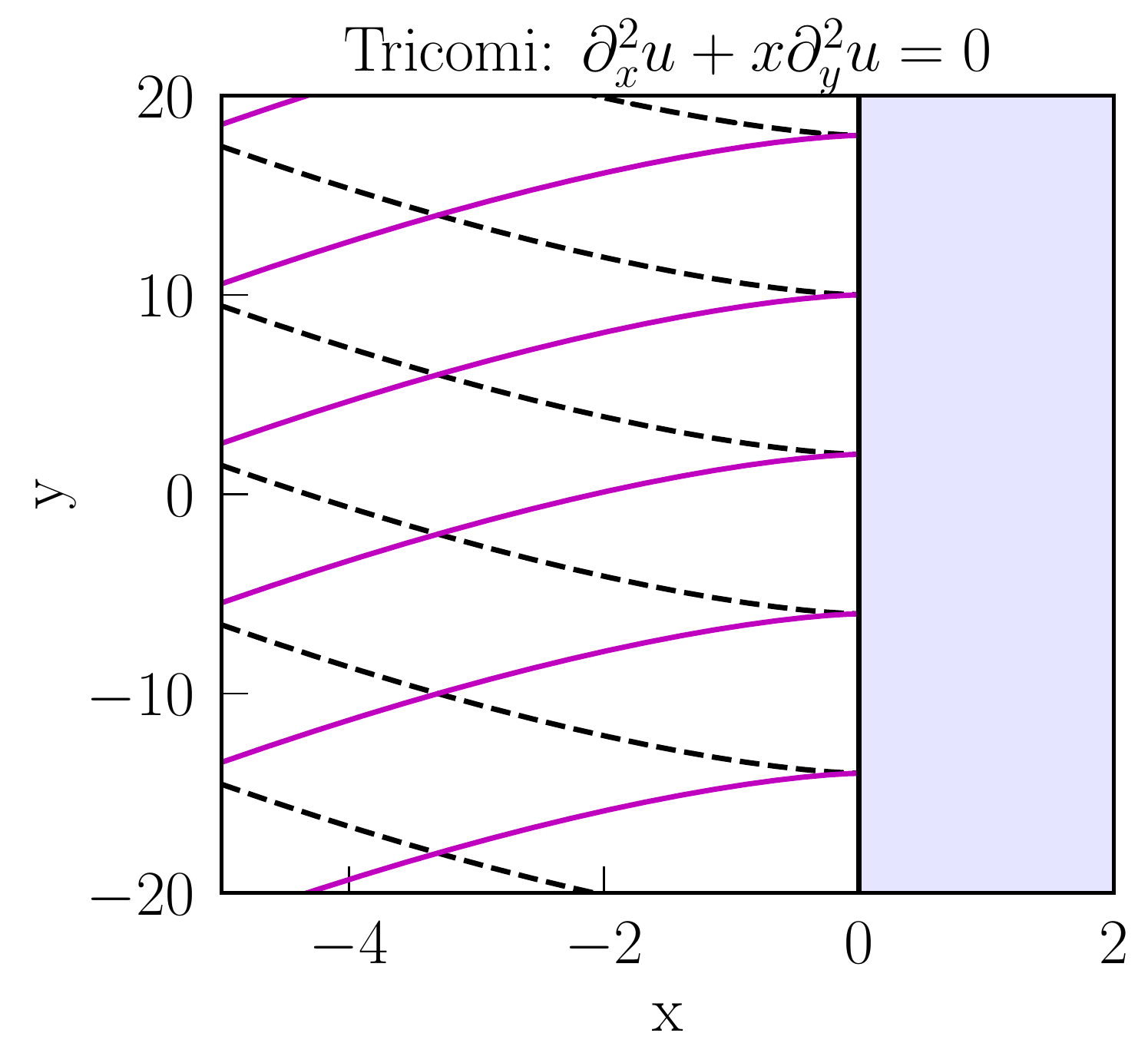}
\includegraphics[width=0.45\textwidth]{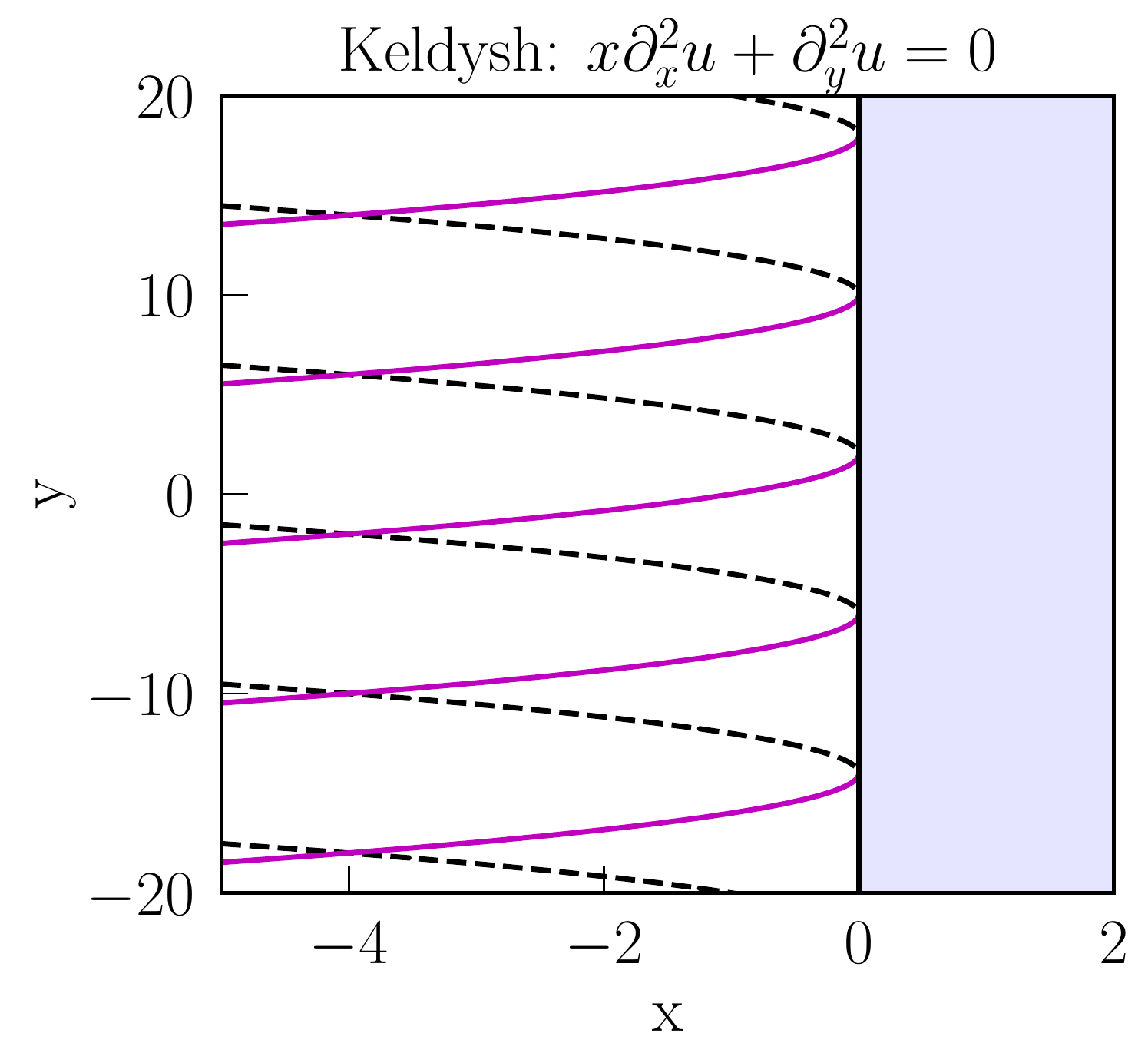}
\caption{The dash black and solid purple lines are the
   integral curves for the characteristics. 
   The elliptic region $x>0$ is shaded in light blue, and the sonic line
   $x=0$ is the solid black vertical line.
   We see that the characteristic
   integral curves for the Tricomi equation intersect the sonic
   perpendicularly, while the characteristic integral curves
   for the Tricomi equation tangentially intersect the sonic line.
   Thus the characteristic speeds blow up before reaching
   the sonic line for the Keldysh equation.
   For more discussion see  
   Morawetz\cite{https://doi.org/10.1002/cpa.3160230404}
   or Otway\cite{otway2015elliptic}.}
   \label{fig:tricomi_keldysh_integral_curves}
\end{figure}

The main qualitative differences between these two equations are how the 
characteristics in the hyperbolic region meet the parabolic sonic line, 
and how the characteristic speeds become imaginary.
Treating $x$ as the ``time'' variable, the characteristic velocities
for the Tricomi and Keldysh equations respectively are
\begin{align}
   c_y^{Tri,\pm}
   =
   \pm\left(-x\right)^{1/2}
   ,\qquad
   c_y^{Kel,\pm}
   =
   \pm\left(-x\right)^{-1/2}
   .
\end{align}
The integral curves of these then are
\begin{align}
   y^{Tri,\pm}
   =
   c \mp \frac{2}{3}\left(-x\right)^{3/2}
   ,\qquad
   y^{Kel,\pm}
   =
   c \mp 2\left(-x\right)^{1/2}
   ,
\end{align}
which we plot in Fig.~\ref{fig:tricomi_keldysh_integral_curves}.
For the Tricomi equation, the characteristics intersect the sonic line
orthogonally, with the corresponding speeds going imaginary passing 
though zero there.
For the Keldysh equation, the characteristics intersect the 
sonic line tangentially, with the characteristic speeds diverging there 
before becoming imaginary .
These properties affect the degree of smoothness one can generally expect for 
solutions to these equations, with the Keldysh equation having weaker 
regularity of solutions on the sonic line\cite{otway2015elliptic}.

\subsection{Systems of PDE in higher dimensions
   \label{sec:systems_higher_dimensions_mixed}
}
We consider systems of first order PDE of the form 
Eq.~\eqref{eq:system_first_order_pde}.
The characteristic equation is (see \ref{sec:review_hyperbolicity}) 
\begin{align}
   \mathrm{det}\left(\mathcal{P}^I_J\left(\xi_{\alpha}\right)\right)
   =
   0
   .
\end{align}
We recall the characteristic equation is defined locally with respect
to a location and solution: 
$\left(x^{\alpha}_0,u^I_0\right)$.
We say that 
the system \eqref{eq:system_first_order_pde} is 
\begin{enumerate}
   \item \emph{hyperbolic} provided the characteristic equation has
all real roots for all possible solutions over the entire domain
   \item \emph{elliptic} if the characteristic equation
has all imaginary roots for all possible solutions over the entire domain.
   \item \emph{mixed-type} PDE if the roots can be real or
imaginary depending on the solution/where in the domain you are
\end{enumerate}
We define the sonic line to define the boundary between elliptic
and hyperbolic domains. There are very few general
results on systems of higher dimensional mixed-type PDE.

\subsection{Example of a mixed-type PDE from Horndeski gravity}

We conclude with an example of a mixed-type system of PDE in a
higher-dimensional spacetime.
We consider k-essence in fixed Minkowski 
spacetime\cite{Armendariz-Picon:2000ulo,
Rendall:2005fv,
Armendariz-Picon:2005oog}, which is simply the
$\mathcal{G}_2$ Horndeski term with no gravity.
The mixed-type properties of this theory were first thoroughly studied by
Bernard et. al.\cite{Bernard:2019fjb} 
(see also Refs.~\refcite{Rendall:2005fv,Akhoury:2011hr,Bezares:2020wkn}). 
Focusing just on the scalar part of the action, we have
\begin{align}
   S_{(\phi)}
   =
   \int d^4\sqrt{-\eta} K\left(X\right)
   .
\end{align}
The equations of motion are
\begin{align}
   \left(
      K' \eta^{\mu\nu}
      -
      K'' \partial^{\mu}\phi\partial^{\nu}\phi
   \right)
   \partial_{\mu}\partial_{\nu}\phi
   =
   0
   .
\end{align}
We define the variables
\begin{align}
   \Phi_{\alpha}
   \equiv
   \partial_{\alpha}\phi
   .
\end{align}
We add to the equations of motion the integrability condition
$\partial_0\Phi_i = \partial_i\Phi_0$.
From this it is straightforward to see that the constraint is propagated in time:
if $C_i \equiv \partial_i\phi - \Phi_i=0$ on a $t=0$ initial data surface, then:
$\partial_0C_i 
= \partial_0\partial_i\phi - \partial_0\Phi_i
= \partial_i\Phi_0 - \partial_0\Phi_i
=
0
$.
The equations of motion in these new variables are 
\begin{align}
   \left[
      \begin{pmatrix}
         K' 
         +
         K'' \Phi_0^2
         \;
         & 
         0
         \\
         0
         &
         \delta^i_j
      \end{pmatrix}
      \partial_0
      -
      \begin{pmatrix}
         2K'' \Phi_0\Phi^j
         \;
         & 
         K'\eta^{ij}
         -
         K'' \Phi^i\Phi^j
         \\
         1
         &
         0
      \end{pmatrix}
      \partial_j
   \right]
   \begin{pmatrix}
      \Phi_0
      \\
      \Phi_i
   \end{pmatrix}
   &=
   0
   .
\end{align}
The principal symbol is 
\begin{align}
   \mathcal{P}
   =
   \begin{pmatrix}
      \left(
         K' 
         +
         K'' \Phi_0^2
      \right)
      \xi_0
      -
      2K'' \Phi_0\Phi^j\xi_j
      \;
      & 
      -
      K'\xi^i
      +
      K'' \Phi^i\Phi^j\xi_j
      \\
      -
      \xi_j
      &
      \delta^i_j\xi_0
   \end{pmatrix}
   .
\end{align}
The determinant is
\begin{align}
   \mathrm{det}\mathcal{P}
   =
   \left[
      \left(
         K' 
         +
         K'' \Phi_0^2
      \right)
      \xi_0^2
      -
      2K'' \Phi_0\Phi^j\xi_j \xi_0
      -
      K'\xi^i\xi_i
      +
      K'' \Phi^i\Phi^j\xi_i\xi_j
   \right]
   \xi_0^2
   .
\end{align}
This has solutions
\begin{align}
   \xi^{(\pm)}_0
   &=
   \frac{1}{K' + K''\Phi_0^2}
   \left(
      K'' \Phi_0\Phi^i\xi_i
      \pm
      \sqrt{
         \left(K' + K''\Phi_0^2\right)
         K'\xi^i\xi_i
      }
   \right)
   ,\\
   \xi_0
   &=
   0
   .
\end{align}
The condition $\xi_0=0$ comes from the integrability condition
$\partial_0\Phi_i=\partial_i\Phi_0$.
We see that the sign of the discriminant to the first equation
determines if the ``physical'' 
(non-constraint and non constraint-violating)
degree of freedom is hyperbolic or elliptic
(we assume $\xi_i$ is real so $\xi^i\xi_i>0$):
\begin{align}
\label{eq:discriminant_k_essence}
   \left(K' + K''\Phi_0^2\right)
   K'
   \begin{cases}
      \geq
      0
      & 
      \mathrm{hyperbolic}
      \\
      <
      0
      &
      \mathrm{elliptic}
   \end{cases}
   .
\end{align}
If $K''=0$, then this is always positive semidefinite. For example
when $K=X$, the theory is always hyperbolic.
Note that the ``constraint'' degrees of freedom are always
hyperbolic (they satisfy $\xi_0=0$).
Studies in spherically symmetric spacetimes,
show that the solutions to $K$-essence theories can take on
Tricomi-like or Keldysh-like character near the formation of
an elliptic region, depending on the specific form of 
$K(X)$\cite{Bernard:2019fjb,Bezares:2020wkn,Lara:2021piy}.
For example, at least in spherically symmetric spacetimes
for couplings $\mathcal{G}_2\left(\phi,X\right)\equiv K\left(X\right)$,
the solutions have Keldysh-like behavior
near sonic lines if\cite{Bezares:2020wkn}
\begin{align}
   1 - 2\frac{K''}{K'}X > 0
   .
\end{align}
Recently work suggests that the
Keldysh-like behavior may be removed through a suitable choice
of gauge\cite{Bezares:2021dma}.
We believe that it is unlikely that elliptic regions 
could be removed for all solutions to all Horndeski theories, 
although there is no rigorous proof that is the case.

We end by noting that in general the ellipticity of the equations
of motion are unrelated to the classical energy conditions. 
For example, in the case of $K$-essence,
the stress-energy tensor is
\begin{align}
   T_{\mu\nu}
   =
   K'\partial_{\mu}\phi\partial_{\nu}\phi
   +
   g_{\mu\nu}
   K
   .
\end{align}
We consider the Null Energy Condition (NEC), which
states that the stress-energy tensor is positive semidefinite
when contracted on all null vectors:
$T_{\mu\nu}k^{\mu}k^{\nu}\geq0,\;\forall k^{\mu} \; s.t. \; k^{\mu}k_{\mu}=0$; 
this condition is used in the proofs of the black hole
area law and singularity theorems\cite{hawking_ellis_1973}.
Letting $k^{\mu}$ be an arbitrary null vector, we have
\begin{align}
   T_{\mu\nu}k^{\mu}k^{\nu}
   =
   K' \left(k^{\mu}\partial_{\mu}\phi\right)^2
   .
\end{align}
Thus the stress-energy tensor obeys the NEC if $K'\geq0$.
We see that this is a distinct condition from the hyperbolicity
of the equations of motion \eqref{eq:discriminant_k_essence}.
For example, if $K=-X$ then the NEC is always violated ($K'=-1$)
(except when $\partial_{\alpha}\phi=0$), but the theory
remains hyperbolic ($\left(K'\right)^2>0$).
While there is no direct relation between the energy
conditions and the classical well-posedness of the Horndeski theories,
it is less clear if the quantized version of
a given NEC violating theory/solution has well-posed evolution. 
For a relatively recent review, see \refcite{Rubakov:2014jja}.

\section{General properties of the principal symbol for Horndeski gravity
   \label{sec:general_properties_principal_symbol}
}
We review some useful properties of the 
principal symbol for the Horndeski theories (or more generally,
theories that have second order equations of motion and that have
a covariant action\cite{Papallo:2017qvl,Kovacs:2020ywu,Reall:2021voz}).
\subsection{Definitions}
The action for Horndeski gravity theories can be written schematically as
\begin{align}
   S
   =
   \int d^4x\sqrt{-g}L\left(g_{\mu\nu},\phi\right)
   .
\end{align}
The equations of motion are
\begin{align}
   \label{eq:eom_general}
   E^{\mu\nu}_{(g)}
   \equiv
   \frac{1}{\sqrt{-g}}\frac{\delta S}{\delta g_{\mu\nu}}
   ,\qquad
   E_{(\phi)}
   \equiv
   \frac{1}{\sqrt{-g}}\frac{\delta S}{\delta\phi}
   .
\end{align}
The principal symbol of the system \eqref{eq:eom_general} is
\begin{align}
\label{eq:definitions_principal_symbol}
   \mathcal{P}\left(\xi\right)
   \begin{pmatrix}
      \delta g_{\rho\sigma} 
      \\
      \delta\phi
   \end{pmatrix}
   \equiv&
   \begin{pmatrix}
      P^{\mu\nu\rho\sigma\alpha\beta}_{(gg)}
      &
      P_{(g\phi)}^{\mu\nu\alpha\beta}
      \\
      P^{\rho\sigma\alpha\beta}_{(\phi g)}
      &
      P^{\alpha\beta}_{(\phi\phi)}
   \end{pmatrix}
   \xi_{\alpha}\xi_{\beta}
   \begin{pmatrix}
      \delta g_{\rho\sigma} 
      \\
      \delta\phi
   \end{pmatrix}
   ,
\end{align}
where
\begin{subequations}
\begin{align}
   &P^{\mu\nu\rho\sigma\alpha\beta}_{(gg)}
   \equiv
   \frac{
      \delta E_{(g)}^{\mu\nu}
   }{
      \delta\left(\partial_{\alpha}\partial_{\beta}g_{\rho\sigma}\right)
   }
   ,\qquad
   &P_{(g\phi)}^{\mu\nu\alpha\beta}
   \equiv
   \frac{
      \delta E_{(g)}^{\mu\nu}
   }{
      \delta\left(\partial_{\alpha}\partial_{\beta}\phi\right)
   }
   ,\nonumber\\
   &P^{\rho\sigma\alpha\beta}_{(\phi g)}
   \equiv
   \frac{
      \delta E_{(\phi)}
   }{
      \delta\left(\partial_{\alpha}\partial_{\beta}g_{\rho\sigma}\right)
   }
   ,
   &P^{\alpha\beta}_{(\phi\phi)}
   \equiv
   \frac{
      \delta E_{(\phi)}
   }{
      \delta\left(\partial_{\alpha}\partial_{\beta}\phi\right)
   }
   .\nonumber
\end{align}
\end{subequations}
\subsection{General symmetries of the principal symbol}
From the definitions in \eqref{eq:definitions_principal_symbol}
we see that
\begin{subequations}
\begin{align}
   P_{(gg)}^{\mu\nu\rho\sigma\alpha\beta}
   =
   P_{(gg)}^{(\mu\nu)\rho\sigma\alpha\beta}
   =
   P_{(gg)}^{\mu\nu(\rho\sigma)\alpha\beta}
   =
   P_{(gg)}^{\mu\nu\rho\sigma(\alpha\beta)}
   ,\\
   P_{(g\phi)}^{\mu\nu\alpha\beta}
   =
   P_{(g\phi)}^{(\mu\nu)\alpha\beta}
   =
   P_{(g\phi)}^{\mu\nu(\alpha\beta)}
   ,\\
   P_{(\phi g)}^{\rho\sigma\alpha\beta}
   =
   P_{(\phi g)}^{(\rho\sigma)\alpha\beta}
   =
   P_{(\phi g)}^{\rho\sigma(\alpha\beta)}
   ,\\
   P_{(\phi\phi)}^{\alpha\beta}
   =
   P_{(\phi\phi)}^{(\alpha\beta)}
   .
\end{align}
\end{subequations}
From the fact that the equations of motion are derived from an action,
the principal symbol is a symmetric matrix\cite{Papallo:2017qvl} 
\begin{subequations}
\begin{align}
   P_{(gg)}^{\mu\nu\rho\sigma\alpha\beta}
   =
   P_{(gg)}^{\rho\sigma\mu\nu\alpha\beta}
   ,\\
   P_{(g\phi)}^{\mu\nu\alpha\beta}
   =
   P_{(\phi g)}^{\mu\nu\alpha\beta}
   .
\end{align}
\end{subequations}

We next consider the additional 
symmetries that come from the action being covariant
(diffeomorphism invariant).
That is, the physical content of the equations of motion are
left invariant by the transformations:
\begin{align}
   \delta g_{\mu\nu}
   \to
   \delta g_{\mu\nu}
   +
   2\nabla_{(\mu}X_{\nu)}
   ,
   \qquad
   \delta\phi
   \to
   \delta\phi
   +
   X^{\mu}\nabla_{\mu}\phi
   ,
\end{align}
where $X^{\mu}$ is a real vector field.
Diffeomorphism invariance implies the Bianchi identity
\begin{align}
   \nabla_{\mu}E^{\mu\nu}_{(g)}
   -
   E_{(\phi)}
   \nabla^{\nu}\phi
   =
   0
   .
\end{align}
To leading order in derivatives this is equal to
\begin{align}
   P_{(gg)}^{\mu\nu\rho\sigma\alpha\beta}
   \partial_{\mu}\partial_{\alpha}\partial_{\beta}g_{\rho\sigma}
   +
   P_{(g\phi)}^{\mu\nu\alpha\beta}
   \partial_{\mu}\partial_{\alpha}\partial_{\beta}\phi
   +
   l.o.t.
   =
   0
   .
\end{align}
Demanding that the highest derivative terms be zero, we have
\begin{subequations}
\begin{align}
   P_{(gg)}^{(\mu|\nu\rho\sigma|\alpha\beta)}
   &=
   0
   ,\\
   P_{(g\phi)}^{(\mu|\nu|\alpha\beta)}
   &=
   0
   .
\end{align}
\end{subequations}
\subsection{The main theorem about the principal symbol
\label{thm:principal_symbol_rewrite}}
Using these symmetries and much rearranging of indices, 
one can prove the following
\begin{enumerate}
\item [] \textbf{Theorem} (Reall\cite{Reall:2021voz}) 
   The principal symbol components can be written as
\begin{align}
   P_{(\phi g)}^{\mu\nu\alpha\beta}
   \xi_{\alpha}\xi_{\beta}
   &=
   C^{\mu\alpha\nu\beta}
   \xi_{\alpha}\xi_{\beta}
   ,\\
   P_{(gg)}^{\mu\nu\rho\sigma\alpha\beta}
   \xi_{\alpha}\xi_{\beta}
   &=
   C^{\mu(\rho|\alpha\nu|\sigma)\beta}\xi_{\alpha}\xi_{\beta}
   ,
\end{align}
where $C^{\mu\nu\rho\sigma}$ has the same symmetries as the Riemann tensor
\begin{align}
\label{eq:riemann_tensor_symmetries}
   C^{\mu\nu\rho\sigma}
   =
   C^{[\mu\nu]\rho\sigma}
   =
   C^{\mu\nu[\rho\sigma]}
   =
   C^{\mu[\nu\rho\sigma]}
   ,
\end{align}
and $C^{\alpha_1\alpha_2\alpha_3\beta_1\beta_2\beta_3}$ has the symmetries
\begin{subequations}
\label{eq:antisymmetric_symmetries}
\begin{align}
   C^{\alpha_1\alpha_2\alpha_3\beta_1\beta_2\beta_3}
   =
   C^{[\alpha_1\alpha_2\alpha_3]\beta_1\beta_2\beta_3}
   =
   C^{\alpha_1\alpha_2\alpha_3[\beta_1\beta_2\beta_3]}
   =
   C^{\beta_1\beta_2\beta_3\alpha_1\alpha_2\alpha_3}
   ,\\
   C^{\alpha_1\alpha_2[\alpha_3\beta_1\beta_2\beta_3]}
   = 
   C^{\alpha_1[\alpha_2\alpha_3\beta_1\beta_2\beta_3]}
   =
   0
   .
\end{align}
\end{subequations}
These ``C-tensors'' are functions of $(g_{\mu\nu},\phi)$ and their
first and second derivatives.
\end{enumerate}
From all the symmetries \eqref{eq:riemann_tensor_symmetries} 
\eqref{eq:antisymmetric_symmetries}, 
we see that we can write
$C^{\alpha_1\alpha_2\alpha_3\beta_1\beta_2\beta_3}$
in terms of a symmetric two-tensor which we call $C_{\mu\nu}$
(this only holds in four spacetime dimensions): 
\begin{align}
   C^{\alpha_1\alpha_2\alpha_3\beta_1\beta_2\beta_3}
   =
   -
   \frac{1}{2}
   \epsilon^{\alpha_1\alpha_2\alpha_3\mu}
   \epsilon^{\beta_1 \beta_2 \beta_3 \nu}
   C_{\mu\nu}
   .
\end{align}
Putting everything together then, the principal symbol for Horndeski
can be written as
\begin{align}
   P\left(\xi\right)
   \begin{pmatrix}
      \delta g_{\rho\sigma}
      \\
      \delta\phi
   \end{pmatrix}
   =
   \begin{pmatrix}
      -
      \frac{1}{2}
      \epsilon^{\mu\rho\alpha\kappa}
      \epsilon^{\nu\sigma\beta\lambda}
      C_{\kappa\lambda}
      &
      C^{\mu\nu\alpha\beta}
      \\
      C^{\mu\nu\alpha\beta}
      &
      P^{\alpha\beta}_{(\phi\phi)}
   \end{pmatrix}
   \xi_{\alpha}\xi_{\beta}
   \begin{pmatrix}
      \delta g_{\rho\sigma}
      \\
      \delta\phi
   \end{pmatrix}
   .
\end{align}
For general relativity (with minimally coupled fields),
$C^{\mu\nu\alpha\beta}=0$ and $C^{\mu\nu}=g^{\mu\nu}$. 
This means for weakly-coupled solutions to the Horndeski theories,
we expect $C^{\mu\nu\alpha\beta}$ to be ``close'' to zero
and $C^{\mu\nu}$ to be ``close'' to $g^{\mu\nu}$.
\subsection{Physical characteristics and the lack of
second time derivatives in the constraint equations
\label{sec:physical_characteristics}}
A real covector $\xi_{\mu}$ is characteristic iff there exists a
nonzero field configuration ${\bf z}\equiv(w_{\mu\nu},v)$ such that
$\mathcal{P}\left(\xi\right){\bf z}=0$.
Due to the diffeomorphism invariance of the equations of motion,
the following field configuration
\begin{align}
   w_{\mu\nu}
   =
   \xi_{(\mu}X_{\nu)}
   ,\qquad
   v
   =
   0
   ,
\end{align}
solves $\mathcal{P}\left(\xi\right){\bf z}=0$ for all $\xi_{\mu}$.
We can ``mod-out'' these unphysical modes by considering equivalence
classes of solutions. 
That is we only consider solutions that are the same
up to the transformation\cite{christodoulou2008mathematical}
\begin{align}
   \left[w_{\mu\nu}\right]
   =
   \left\{
      t_{\mu\nu}
      :
      \;
      \exists \xi_{\mu}
      \;
      \;
      s.t.
      \;
      \;
      t_{\mu\nu}
      =
      w_{\mu\nu}
      +
      \xi_{(\mu}X_{\nu)}
   \right\}
.
\end{align}
We can then think of the principal symbol as having 
the physical eigenvectors
$\left[{\bf z}\right]=\left(\left[w_{\mu\nu}\right],v\right)$, .

We next present a more general argument that shows
the constraint equations have no second time derivatives
acting on the fields in adapted ADM coordinates.
As was shown by Reall\cite{Reall:2021voz} 
(see \ref{thm:principal_symbol_rewrite}),
the principal part of the tensor equations of motion can be written as
\begin{align}
   \left(E^{(g)}\right)^{\mu\nu}
   =
   C^{\mu(\rho|\alpha\nu|\sigma)\beta}
   \partial_{\alpha}\partial_{\beta}g_{\rho\sigma}
   +
   C^{\mu\alpha\nu\beta}
   \partial_{\alpha}\partial_{\beta}\phi_I
   +
   l.o.t.
   ,
\end{align}
where $C^{\mu\alpha\nu\beta}$ has the same symmetries as the Riemann
tensor, and $C^{\mu(\rho|\alpha\nu|\sigma)\beta}$ has the following symmetries:
\begin{subequations}
\begin{align}
   C^{\alpha_1\alpha_2\alpha_3\beta_1\beta_2\beta_3}
   =
   C^{[\alpha_1\alpha_2\alpha_3]\beta_1\beta_2\beta_3}
   =
   C^{\alpha_1\alpha_2\alpha_3[\beta_1\beta_2\beta_3]}
   =
   C^{\beta_1\beta_2\beta_3\alpha_1\alpha_2\alpha_3}
   ,\\
   C^{\alpha_1\alpha_2[\alpha_3\beta_1\beta_2\beta_3]}
   = 
   C^{\alpha_1[\alpha_2\alpha_3\beta_1\beta_2\beta_3]}
   =
   0
   .
\end{align}
\end{subequations}
In ADM coordinates we have $n_{\mu}=-N\delta_{\mu}^0$ (where $N$ is the lapse;
see Eq.~\eqref{eq:3p1_decomposition}), so that 
\begin{subequations}
\begin{align}
   \mathcal{H}
   &=
   n_{\mu}n_{\nu}
   \left(
      C^{\mu(\rho|\alpha\nu|\sigma)\beta}
      \partial_{\alpha}\partial_{\beta}g_{\rho\sigma}
      +
      C^{\mu\alpha\nu\beta}
      \partial_{\alpha}\partial_{\beta}\phi
   \right)
   +
   l.o.t.
   ,\nonumber\\
   &=
   N^2
   \left(
      C^{0(\rho|\alpha0|\sigma)\beta}
      \partial_{\alpha}\partial_{\beta}g_{\rho\sigma}
      +
      C^{0\alpha0\beta}
      \partial_{\alpha}\partial_{\beta}\phi
   \right)
   +
   l.o.t.
   ,\\ 
   \mathcal{M}_{\mu}
   &=
   n_{\mu}h_{\nu\gamma}
   \left(
      C^{\mu(\rho|\alpha\gamma|\sigma)\beta}
      \partial_{\alpha}\partial_{\beta}g_{\rho\sigma}
      +
      C^{\mu\alpha\nu\beta}
      \partial_{\alpha}\partial_{\beta}\phi
   \right)
   +
   l.o.t.
   \nonumber\\
   &=
   -Nh_{\nu\gamma}
   \left(
      C^{0(\rho|\alpha\gamma|\sigma)\beta}
      \partial_{\alpha}\partial_{\beta}g_{\rho\sigma}
      +
      C^{0\alpha\nu\beta}
      \partial_{\alpha}\partial_{\beta}\phi
   \right)
   +
   l.o.t.
   .
\end{align}
\end{subequations}
From the symmetries of the tensors
$C^{\alpha_1\alpha_2\beta_1\beta_2}$ and 
$C^{\alpha_1\alpha_2\alpha_3\beta_1\beta_2\beta_3}$,
we see that the terms with $\partial_0^2$ acting on the tensor
or scalar field are zero.

\section{Identities for the $3+1$ decomposition of spacetime, and 
conformal transverse-traceless decomposition
   \label{sec:spatial_conformal_decomposition}
}
Here we collect a few formulas that are useful for the $1+3$ decomposition
of the equations of motion, and the conformal transverse-traceless
decomposition of the constraint equations.
We refer to Sec.~\ref{sec:constructing_initial_data} for definitions.

The nonzero contractions of the Riemann tensor 
are\cite{Cook:2000vr,baumgarte2010numerical,Gourgoulhon:2007ue}
\begin{subequations}
\begin{align}
   h_{\alpha_1}^{\mu_1}
   h_{\alpha_2}^{\mu_2}
   h_{\alpha_3}^{\mu_3}
   h_{\alpha_4}^{\mu_4}
   R_{\mu_1\mu_2\mu_3\mu_4}
   =&
   {}^{(3)}R_{\alpha_1\alpha_2\alpha_3\alpha_4}
   +
   K_{\alpha_1\alpha_3}K_{\alpha_2\alpha_4}
   -
   K_{\alpha_1\alpha_4}K_{\alpha_2\alpha_3}
   ,\\
   h_{\alpha_1}^{\mu_1}
   h_{\alpha_2}^{\mu_2}
   h_{\alpha_3}^{\mu_3}
   n^{\mu_4}
   R_{\mu_1\mu_2\mu_3\mu_4}
   =&
   D_{\alpha_2}K_{\alpha_1\alpha_3}
   -
   D_{\alpha_1}K_{\alpha_2\alpha_3}
   ,\\
   h_{\alpha_1}^{\mu_1}
   n^{\mu_2}
   h_{\alpha_3}^{\mu_3}
   n^{\mu_4}
   R_{\mu_1\mu_2\mu_3\mu_4}
   =&
   \mathcal{L}_{\bf n}K_{\alpha_1\alpha_3}
   +
   \frac{1}{N}D_{\alpha_1}D_{\alpha_3}N
   +
   K_{\alpha_1}^{\gamma}K_{\alpha_2\gamma}
   .
\end{align}
\end{subequations}
Similarly the nonzero contractions on the scalar field second derivatives are 
\begin{align}
   h_{\mu}{}^{\gamma}h^{\nu}{}_{\delta}\nabla_{\gamma}\nabla^{\delta}\phi
   &=
   D_{\mu}D^{\nu}\phi
   +
   n^{\nu}K_{\mu}^{\delta}D_{\delta}\phi
   +
   K_{\mu}{}^{\nu}\mathcal{L}_{\bf n}\phi
   ,\\
   h_{\mu}{}^{\gamma}n_{\delta}\nabla_{\gamma}\nabla^{\delta}\phi
   &=
   D_{\mu}\left(\mathcal{L}_{\bf n}\phi\right)
   +
   K_{\mu}^{\delta}D_{\delta}\phi
   ,\\
   n^{\gamma}n_{\delta}\nabla_{\gamma}\nabla^{\delta}\phi
   &=
   \mathcal{L}_{\bf n}\mathcal{L}_{\bf n}\phi
   -
   \left(n^{\gamma}\nabla_{\gamma}n^{\delta}\right)\nabla_{\delta}\phi
   ,
\end{align}
Where $\mathcal{L}_{\bf n}$ denotes the Lie derivative along $n^{\mu}$.

We next list a few useful formulas for the CTT 
decomposition\cite{Gourgoulhon:2007ue,Kovacs:2021lgk}.
The CTT decomposition rewrites the spatial metric as 
\begin{align}
   h_{ij}
   &=
   \psi^4\tilde{h}_{ij}
   .
\end{align}
We denote the metric compatible derivative with respect to $h_{ij}$
as $D_i$, and the metric compatible derivative with respect to
$\tilde{h}_{ij}$ as $\tilde{D}_i$.
We similarly denote the Christoffel symbols and Riemann
tensor with respect to each metric. We have 
\begin{subequations}
\begin{align}
   {}^{(3)}\Gamma^k_{ij}
   =&
   {}^{(3)}\tilde{\Gamma}^k_{ij}
   +
   C^k_{ij}
   \nonumber\\
   \equiv&
   {}^{(3)}\tilde{\Gamma}^k_{ij}
   +
   2
   \left(
      \delta^k_i\tilde{D}_j\ln\psi
      +
      \delta^k_j\tilde{D}_i\ln\psi
      -
      \tilde{h}_{ij}\tilde{D}^k\ln\psi
   \right)
   \\
   {}^{(3)}R^{i_1i_2}{}_{j_1j_2}
   =&
   \frac{1}{\psi^4}{}^{(3)}\tilde{R}^{i_1i_2}{}_{j_1j_2} 
   +
   2\frac{1}{\psi^4}\tilde{h}^{i_2k}\tilde{D}_{[j_1}C^{i_1}_{j_2]k}
   +
   2\frac{1}{\psi^4}\tilde{h}^{i_2k}C^{i_2}_{l[j_1}C^l_{j_2]k}
   \nonumber\\
   =&
   \frac{1}{\psi^4}{}^{(3)}\tilde{R}^{i_1i_2}{}_{j_1j_2} 
   -
   8
   \frac{1}{\psi^5}
   \delta^{[i_1}_{[j_1}\tilde{D}_{j_2]}\tilde{D}^{i_2]}\psi
   \nonumber\\
   &+
   24\frac{1}{\psi^6}\delta^{[i_1}_{[j_1}
   \tilde{D}_{i_2]}\psi\tilde{D}^{i_2]}\psi
   -
   4\delta^{i_1i_2}_{j_1j_2}\tilde{D}_k\psi\tilde{D}^k\psi
   .
\end{align}
\end{subequations}
\section{$4\partial ST$ gravity as a gradient expansion about 
   General Relativity with a scalar field
   \label{sec:4dst_gradient_expansion}
}
   Here we review an effective field theory/gradient expansion-styled 
argument to motivate $4\partial ST$ 
gravity\cite{Weinberg:2008hq,Kovacs:2020pns,Kovacs:2020ywu}.
We assume a zeroeth order action
\begin{align}
\label{eq:lead_order_4dst}
   S_0
   =
   \int d^4x\sqrt{-g}\left(
      \frac{1}{2}R
      +
      X
      -
      V\left(\phi\right)
   \right)
   .
\end{align}
Assuming a gradient expansion about this action,
the leading order corrections to \eqref{eq:lead_order_4dst}
will contain all of the covariant terms with four spacetime derivatives
and order-unity coefficients (there can be no third covariant terms
with only three derivatives in a $4D$ spacetime).
Up to total derivatives and integration by parts, these are
\begin{align}
\label{eq:all_next_terms_4dst}
   S_{\partial^4}
   =
   \int d^4x\sqrt{-g}\Big(
      &
      f_1\left(\phi\right)X^2
      +
      f_2\left(\phi\right)X\Box\phi
      +
      f_3\left(\phi\right)\left(\Box\phi\right)^2
      \nonumber\\
      &+
      f_4\left(\phi\right)R^{\mu\nu}\nabla_{\mu}\phi\nabla_{\nu}\phi
      +
      f_5\left(\phi\right)RX
      +
      f_6\left(\phi\right)R\Box\phi
      \nonumber\\
      &+
      f_7\left(\phi\right)R^2
      +
      f_8\left(\phi\right)R^{\mu\nu}R_{\mu\nu}
      \nonumber\\
      &+
      f_9\left(\phi\right)\mathcal{R}_{GB}
      +
      f_{10}\left(\phi\right)\mathcal{R}_{P}
   \Big)
   .
\end{align}
The four dimensional Gauss-Bonnet scalar $\mathcal{R}_{GB}$ and the
Pontryagin density $\mathcal{R}_{P}$ are
\begin{align}
   \mathcal{R}_{GB}
   \equiv&
   \frac{1}{4}\delta^{\mu\nu\rho\sigma}_{\alpha\beta\gamma\delta}
   R^{\alpha\beta}{}_{\mu\nu}
   R^{\gamma\delta}{}_{\rho\sigma}
   ,\\ 
   \mathcal{R}_{P}
   \equiv&
   \frac{1}{2}
   \epsilon^{\gamma\delta\mu\nu}
   R^{\alpha\beta}{}_{\mu\nu}
   R_{\beta\alpha\gamma\delta}
   ,
\end{align}
where $\delta^{\cdots}_{\cdots}$ is the generalized Kronecker delta
tensor and $\epsilon^{\mu\nu\gamma\delta}$ is the Levi-Cevita tensor.
Besides using $X$ instead of $\left(\nabla\phi\right)^2$, the action
\eqref{eq:all_next_terms_4dst} differs from Weinberg's\cite{Weinberg:2008hq}
Eq.~(3) in that we use the Gauss-Bonnet scalar $\mathcal{R}_{GB}$ instead
of the contraction of the Weyl tensor:
$C_{\mu\nu\alpha\beta}C^{\mu\nu\alpha\beta}$
(the Pontryagin density can be written equivalently in terms
of contractions of the Weyl tensor\cite{Grumiller:2007rv}).
Up to field redefinitions and total derivatives, 
these two terms can be interchanged with one 
another\cite{Kovacs:2020pns,Kovacs:2020ywu}.

If we were to consider the equations of motion in
Eq.~\eqref{eq:all_next_terms_4dst}, the equations of motion would
be fourth order in time for both the tensor and scalar field, 
which would imply there are two new degrees of freedom.
We must remove these two ``spurious'' degrees of freedom 
in order for the solution to be consistent with the gradient expansion.
This is accomplished by using the lower order equations of motion
to remove higher derivative terms. This can be done several ways:
\begin{enumerate}
   \item By removing higher derivative terms directly in the action by
using the lower-order equations of motion.
   \item By removing the higher derivative terms in the equations of motion
      by making use of the lower-order equations of motion.
   \item By solving the full system of equations of motion order-by-order
      in a small expansion parameter (order-reduction methods).
\end{enumerate}
Following Weinberg, we focus on the first approach.
The lower-order equations of motion are found by varying
Eq.~\eqref{eq:lead_order_4dst} to obtain:
\begin{align}
   R_{\mu\nu}
   -
   \nabla_{\mu}\phi\nabla_{\nu}\phi
   -
   g_{\mu\nu}V
   &=
   0
   ,\\
   \Box\phi - V'
   &=
   0
   .
\end{align}
With these equations we can remove all higher derivative
terms that have powers of $R_{\mu\nu},R$ and $\Box\phi$.
Relabeling things, we see that the final action can be written as
\begin{align}
   S
   &=
   \int d^4x\sqrt{-g}\left(
      \frac{1}{2}R
      +
      X
      -
      V\left(\phi\right)
      +
      \alpha\left(\phi\right)X^2
      +
      \beta\left(\phi\right)\mathcal{R}_{GB}
      +
      \gamma\left(\phi\right)\mathcal{R}_{P}
   \right)
   .
\end{align}
The Pontryagin density is not invariant under parity ${\bf x}\to-{\bf x}$;
if $\gamma\neq0$ then we would need $\phi$ to be a pseudoscalar and the
couplings to take a particular form if we wanted the action to be
invariant under parity\cite{Alexander:2009tp}.
If we assume $\phi$ is a scalar, and that the action is invariant under
parity transforms, we are left with $4\partial ST$ gravity:
\begin{align}
   S_{4\partial ST}
   &=
   \int d^4x\sqrt{-g}\left(
      \frac{1}{2}R
      +
      X
      -
      V\left(\phi\right)
      +
      \alpha\left(\phi\right)X^2
      +
      \beta\left(\phi\right)\mathcal{R}_{GB}
   \right)
   .
\end{align}
This action has second order equations of motion, so the higher derivative
terms do not induce new degrees of freedom.
Thus in $4\partial ST$ gravity we can solve the full equations of motion
and obtain a solution consistent with the gradient expansion,
provided the higher derivative terms make a suitable ``small'' contribution
to the solution\cite{Kovacs:2020pns,Kovacs:2020ywu}.
If we had kept the Pontryagin term, the equations of motion
would not be second order\cite{Delsate:2014hba}.
Order-reductions methods are currently being developed for application
in numerical relativity to solve for the dynamics of theories that
have the scalar field-Pontryagin coupling\cite{Okounkova:2017yby,
Okounkova:2018abo,Okounkova:2018pql,GalvezGhersi:2021sxs
}.

\bibliography{thebib}
\bibliographystyle{hunsrt}

\end{document}